\documentclass{natureJLH}

\usepackage{graphicx}
\usepackage[none]{hyphenat}
\usepackage{tikz}
\usepackage[normalem]{ulem} % use the option normalem; otherwise the journal titles will be under lined  
\usepackage{lineno}
\usepackage{amssymb}
\usepackage{amsmath}
\usepackage{caption}
\usepackage{booktabs,multirow}
\usepackage{multirow}
\usepackage{pdflscape}
\usepackage{xcolor}
\usepackage{hyperref}
\usepackage{pifont}
\usepackage{soul}
\usepackage{amsmath}
\usepackage{pdfpages}
\newcommand{\araa}{Annu. Rev. Astron. Astrophys.}  % Annual Review of Astron and Astrophys
\newcommand{\aj}{Astron. J.}   % Astronomical Journal
\newcommand{\apj}{Astrophys. J.}   % Astrophysical Journal
\newcommand{\apjs}{Astrophys. J. Suppl. Ser.}   % Astrophysical Journal, Supplement
\newcommand{\aap}{Astron. Astrophys.}   % Astronomy and Astrophysics
\newcommand{\mnras}{Mon. Not. R. Astron. Soc.}   % Monthly Notices of the RAS
\newcommand{\nat}{Nature} % Nature
\newcommand{\nastro}{Nat. Astron.} % Nature Astronomy
\newcommand{\na}{New Astron.}   % New Astronomy
\newcommand{\pasp}{Publ. Astron. Soc. Pac.}   % Publications of the Astron. Soc. of the Pacific
\newcommand{\raa}{Res. Astron. Astrophys.} % Research in Astronomy and Astrophysics (formerly CJAA)
\newcommand{\sci}{Science} % Science
\title{A potential mass-gap black hole in a wide binary with a circular orbit}

\author{Song Wang$^{\star1,2,20,21}$, Xinlin Zhao$^{1,3,20}$,  Fabo Feng$^{\star4,5,21}$, Hongwei Ge$^{6,7,8}$, Yong Shao$^{9,10}$, Yingzhen Cui$^{1}$, Shijie Gao$^{9,10}$, Lifu Zhang$^{6,3,8}$, Pei Wang$^{11,2}$, Xue Li$^{1,3}$, Zhongrui Bai$^{1}$, Hailong Yuan$^{1}$, Yang Huang$^{3,1}$, Haibo Yuan$^{12,2}$, Zhixiang Zhang$^{13}$, Tuan Yi$^{14,15}$, Maosheng Xiang$^{1,2}$, Zhenwei Li$^{6, 8}$, Tanda Li$^{12,2}$, Junbo Zhang$^{1}$, Meng Zhang$^{1}$, Henggeng Han$^{1}$, Dongwei Fan$^{1}$, Xiangdong Li$^{9,10}$, Xuefei Chen$^{6,7,8}$, Zhengwei Liu$^{6,7,8}$, Xiangcun Meng$^{6,8,16}$, Qingzhong Liu$^{17,18}$, Haotong Zhang$^{1}$, Wei-Min Gu$^{13}$, and Jifeng Liu$^{\star1,2,3,19,21}$. (\today)}

\begin{document}

\maketitle

\begin{affiliations}

\item Key Laboratory of Optical Astronomy, National Astronomical Observatories, Chinese Academy of Sciences, Beijing 100101, China
\item Institute for Frontiers in Astronomy and Astrophysics, Beijing Normal University, Beijing 102206, China
\item School of Astronomy and Space Sciences, University of Chinese Academy of Sciences, Beijing 100049, China
\item Tsung-Dao Lee Institute, Shanghai Jiao Tong University, Shanghai, 201210, China
\item School of Physics and Astronomy, Shanghai Jiao Tong University, Shanghai 200240, China
\item Yunnan Observatories, Chinese Academy of Sciences, Kunming, 650216, China
\item International Centre of Supernovae, Yunnan Key Laboratory, Kunming 650216, China
\item Key Laboratory for Structure and Evolution of Celestial Objects, Chinese Academy of Sciences, Kunming 650216, China
\item School of Astronomy and Space Science, Nanjing University, Nanjing 210023, China
\item Key Laboratory of Modern Astronomy and Astrophysics (Nanjing University), Ministry of Education, Nanjing 210023, China
\item CAS Key Laboratory of FAST, NAOC, Chinese Academy of Sciences, Beijing 100101, China
\item Department of Astronomy, Beijing Normal University, Beijing 100875, China
\item Department of Astronomy, Xiamen University, Xiamen, Fujian 361005, China
\item Department of Astronomy, School of Physics, Peking University, Beijing 100871, China
\item Kavli Institute for Astronomy and Astrophysics, Peking University, Beijing 100871, China
\item Center for Astronomical Mega-Science, Chinese Academy of Sciences, Beijing 100012, China
\item Key Laboratory of Dark Matter and Space Astronomy, Purple Mountain Observatory, Chinese Academy of Sciences, Nanjing 210008, China
\item College of Physics and Electronic Engineering, Qilu Normal University, Jinan, China
\item New Cornerstone Science Laboratory, National Astronomical Observatories, Chinese Academy of Sciences, Beijing 100012, China
\item These authors contributed equally: Song Wang, Xinlin Zhao.
\item Corresponding authors:  songw@bao.ac.cn; ffeng@sjtu.edu.cn; jfliu@nao.cas.cn

\end{affiliations}

\begin{abstract}

Mass distribution of black holes identified through X-ray emission suggests a paucity of black holes in the mass range of 3 to 5 solar masses.
Modified theories have been devised to explain this mass gap, and it is suggested that natal kicks during supernova explosion can more easily disrupt binaries with lower mass black holes.
Although recent LIGO observations reveal the existence of compact remnants within this mass gap, the question of whether low-mass black holes can exist in binaries remains a matter of debate.
Such a system is expected to be noninteracting without X-ray emission, and can be searched for using radial velocity and astrometric methods.
Here we report Gaia DR3 3425577610762832384, a wide binary system including a red giant star and an unseen object, exhibiting an orbital period of approximately 880 days and near-zero eccentricity.
Through the combination of radial velocity measurements from LAMOST and astrometric data from Gaia DR2 and DR3 catalogs, we determine a mass of $3.6^{+0.8}_{-0.5}$ $M_{\odot}$ of the unseen component.
This places the unseen companion within the mass gap, strongly suggesting the existence of binary systems containing low-mass black holes.
More notably, the formation of its surprisingly wide circular orbit challenges current binary evolution and supernova explosion theories.

\end{abstract}

Using spectroscopy obtained from the Large Sky Area Multi-Object Fiber Spectroscopic Telescope (LAMOST)\cite{2012RAA....12.1197C} and astrometry data from Gaia\cite{GaiaCollaboration2016}, we conducted a search for stars exhibiting radial velocity (RV) variation and astrometric solutions, aiming at identifying binaries with compact components.
We discovered a most promising black hole candidate in the mass gap, which was defined by the scarcity of black holes with masses ranging from 3 to 5 $M_{\odot}$\cite{1998ApJ...499..367B}.
The black hole is part of a binary system with the Gaia ID 3425577610762832384 (hereafter G3425), positioned at coordinates (R.A., decl.) = (94.27876$^{\rm o}$, 23.73022$^{\rm o}$).

The motion of the visible star in the binary can be traced through 27 LAMOST low- and medium-resolution observations spanning seven years.
By fitting a binary orbit to the RV data, we derived a period of $P$ = 877$\pm$2 days, a center-of-mass velocity $V_{\rm 0}$ = $-$10.7$\pm$0.2 km s$^{-1}$, a semi-amplitude $K_{\rm 1}$ $=$ 22.9$\pm$0.1 km s$^{-1}$, and an eccentricity $e$ $=$ 0.05$\pm$0.01, consistent with the Gaia solution in the {\sc nss\_two\_body\_orbit} table\cite{2023A&A...674A..34G} (Extended Data Fig. 1).
The LAMOST medium-resolution spectra exhibit only a single set of absorption lines, indicating G3425 is a single-lined spectroscopic binary including an unseen component.
The binary mass function can be calculated as $f(M) = \frac{M_{\rm 2} {\rm sin}^3 i} {(1+q)^{2}} = \frac{P K_{1}^{3} \, (1-e^2)^{3/2}}{2\pi G} = 1.09\pm 0.02\ M_{\rm \odot}$ (Extended Data Table 1), where $q=M_1/M_2$, $M_1$ is the mass of the visible star and $M_2$ is the mass of the unseen object.
$G$ is the gravitational constant and $i$ is the orbital inclination angle.

The distance of the system can be derived from Gaia parallax only when considering it as a wide binary (including an invisible component), while a single-star astrometric solution may lead to inaccuracies.
Take Gaia BH2 as an example\cite{2023MNRAS.521.4323E}, the parallax values differ between Gaia data release (DR) 2, DR3, and the {\sc nss\_two\_body\_orbit} (NTBO) catalog, which are $1.58\pm0.03$, $0.67\pm0.10$, and $0.86\pm0.02$ mas, respectively.
Similarly, for G3425, Gaia provides different parallax values, with DR2 reporting $0.04\pm0.05$ mas and DR3 reporting $0.70\pm0.05$ mas, corresponding to a distance of 6925$\pm$1670 pc and 1414$\pm$82 pc, respectively.
To address these inconsistencies, we conducted a joint analysis using the RV data from LAMOST and astrometric data from both Gaia DR2 and DR3, and used the adaptive Markov Chain Monte Carlo (MCMC) method to sample the posterior of the RV and astrometric models\cite{2019ApJS..242...25F} (Methods and Extended Data Figs. 2--3). 
This allowed us to simultaneously determine the astrometry of the binary barycenter and orbital parameters.
As a result, we derived a new parallax estimate of $0.56^{+0.09}_{-0.09}$ mas, corresponding to a distance of 1786$^{+342}_{-248}$ pc, and an inclination angle of $i = 89^{+15}_{-10}$ degrees, suggesting G3425 is a near edge-on system (Extended Data Table 2).

The stellar parameters of the visible star are consistent with a giant star (Extended Data Fig. 4), supported by both spectroscopic and spectral energy distribution (SED) analysis (Supplementary Information).
Based on the atmospheric parameters obtained from LAMOST DR9, G3425 has an effective temperature of $T_{\rm eff} = 4984\pm 25$ K, a surface gravity of ${\rm log}g = 2.63\pm 0.05$, and a metallicity of [Fe/H] $=-0.12\pm 0.02$.
The extinction, $E(B-V)$, estimated from the StarPair method\cite{2013MNRAS.430.2188Y} is 0.46$\pm$0.01, slightly higher than the value derived from the 3D dust map\cite{2019ApJ...887...93G} along G3425's direction ($E(B-V) = 0.37^{+0.04}_{-0.02}$).
SED fitting (Extended Data Fig. 5) returns consistent parameters with the spectroscopic analysis, including $T_{\rm eff} = 4990^{+25}_{-21}$ K, ${\rm log}g = 2.64^{+0.04}_{-0.05}$, [Fe/H]$=-0.13^{+0.02}_{-0.01}$, and $E(B-V) = 0.45^{+0.01}_{-0.01}$.
The abundances of 15 elements (e.g., $\alpha$, Mg, Al) show no anomalies for a red giant star.

The mass of the unseen component can be obtained using the mass of the giant star and orbital parameters.
Utilizing multi-band magnitudes, together with the new distance estimation and atmospheric parameters, we calculated the spectroscopic mass of the giant, following $M = gL_{\rm bol}/(4\pi G\sigma T_{\rm eff}^{4})$ where $L_{\rm bol}$ is the bolometric luminosity and $\sigma$ is the Stefan-Boltzmann constant, to be $M_{\rm 1} = 2.66^{+1.18}_{-0.68}$ $M_\odot$ and the radius to be $R_{\rm 1} = 12.97_{-1.77}^{+2.43}$ $R_\odot$.
As a check, employing single-star evolutionary models, we derived the stellar mass, radius, and age of the giant to be $M_{\rm 1} = 2.39^{+0.22}_{-0.18}$ $M_{\odot}$, $R_{\rm 1} = 12.54_{-1.02}^{+0.85}$ $R_\odot$, and $t = 0.79^{+0.18}_{-0.21}$\,Gyr, consistent with the spectroscopic analysis.
Given the binary mass function and the spectroscopic mass of the visible star, along with the inclination angle estimate, we calculated the mass of the unseen object to be $3.6^{+0.8}_{-0.5}$ $M_{\odot}$, exceeding the upper limit mass for a neutron star ($\lesssim$3 $M_{\odot}$).
The atmospheric parameters estimated from different methods can result in different spectroscopic mass values of the giant ranging from 1.7 $M_{\odot}$ to 2.9 $M_{\odot}$, which led to a mass of the unseen object falling within a range of 2.9 $M_{\odot}$ to 3.7 $M_{\odot}$.
Neither pulsed nor persistent emission was detected from radio observations with the Five-hundred-meter Aperture Spherical radio Telescope\cite{2020RAA....20...64J} (Supplementary Information).

It's necessary to consider whether the giant in the G3425 system might be a stripped star with ongoing mass transfer, as suggested for V723 Mon\cite{2022MNRAS.512.5620E} and 2M0412\cite{2022MNRAS.516.5945J}, or with post mass transfer, such as HR 6819\cite{2021MNRAS.502.3436E} (Methods). In such cases, the giant may have a significantly lower mass than predicted by single-star evolutionary models.
Here we present evidence against these possibilities.
The radius of the visible star of G3425 is approximately 13 $R_{\odot}$, representing only about 4.5\% of the Roche lobe radius ($\sim$ 290 $R_\odot$), suggesting it is far from filling its Roche lobe. 
If the giant has been stripped and has a current mass of 0.5--2.0 $M_{\odot}$, the inferred mass of the companion is about 1.79--3.02 $M_{\odot}$, corresponding to a luminosity ratio of 0.12--0.67 compared to the giant assuming the companion is a main-sequence star.
However, comparisons between G3425's spectrum and giants with similar atmospheric parameters show quite good agreement (Fig. 1), revealing no excess absorption or emission in spectral lines from other components.
Simultaneously, a stripped star (post-mass transfer) with such a low ratio of radius to Roche lobe radius ($\approx$4.5\%) typically would have a surface temperature exceeding 10,000 $K$\cite{2017MNRAS.467.1874C,2018A&A...615A..78G}.
Most importantly, through spectral disentangling, it is confirmed that there is no light contribution from any other component in the system (Extended Data Fig. 6).
Additionally, a search of evolutionary paths using the Binary Population and Spectral Synthesis code finds no post-interaction binary model with similar properties to G3425.

The nature of the unseen component cannot be one or a pair of main-sequence stars, as proposed for the black hole candidate 2M05215658+4359220\cite{2019Sci...366..637T,2020Sci...368.3282V}.
Assuming a main-sequence star with a mass of 3.6 $M_{\odot}$, its $G$-band absolute magnitude is about $-$0.2 mag, comparable to the luminosity of the visible star in G3425.
However, no flux excess in the blue band is observed in the flux-calibrated Gaia XP spectrum of G3425, compared with a spectral template (Fig. 1).
On the other hand, if the companion is a binary consisting of two stars with masses of $\approx$1.8 $M_{\odot}$, the $G$- and $BP$-band luminosity ratio between the close binary and the visible giant are approximately 1/4 and 1/3, respectively.
Such an inner binary can be easily detectable in the blue part of the spectra or SED of G3425, thus safely excluding this scenario.
These make G3425 a highly promising low-mass black hole falling within the mass gap (Fig. 2).

The formation of G3425 challenges our understanding of the processes of binary evolution and supernova explosion.
Similar to Gaia BH1 and BH2, its orbit appears too wide to have formed through common envelope evolution\cite{2023MNRAS.518.1057E,2023MNRAS.521.4323E}.
Our numerical simulation using an adiabatic mass-loss model\cite{2010ApJ...717..724G,2022ApJ...933..137G,2023ApJ...945....7G} (Supplementary Information) showed that a wide orbit similar to G3425 could only be produced by adopting a high ejection efficiency of common envelope (i.e., $\alpha_{\rm CE} = $ 10).
Moreover, G3425 exhibits a surprisingly circular orbit in contrast to Gaia BH1 and BH2 (Fig. 2).
The lower mass of the black hole in G3425 indicates more material needs to be ejected, making the binary more likely to be unbound or have a high eccentricity, which cannot be circularized by tidal torque within a Hubble time.
Clearly, G3425 also cannot form through a dynamical capture of the giant star by a black hole.
Our binary population synthesis, using the Binary Star Evolution (BSE) code, showed that the stochastic prescription of supernova-explosion mechanism\cite{2020MNRAS.499.3214M}, which involves relatively small natal kicks, can generate systems like G3425, although such systems are exceptionally rare, with a number ranging from $\approx$0.02 to $\approx$1 in the Milky Way corresponding to common envelope ejection efficiencies varying from 1 to 5 (Methods and Extended Data Fig. 7). 

An alternative is that G3425 was originally a triple system, with the observed giant star as an outermost component and an inner binary containing two massive stars (Supplementary Information).
The present black hole formed as a result of a merger of the inner binary after long-term evolution.
It is also possible that the central unseen object still contains two less-massive compact objects (e.g., two neutron stars, or a neutron star and a massive white dwarf).
In this case, it would be an exciting candidate of the merger of binary neutron stars or a neutron star and a white dwarf, which can be detected in the future by gravitational wave observations. 
More precise short-cadence spectroscopic monitoring is required to search for any short-period RV modulation caused by an inner compact binary, although it is quite difficult to vet.

The rare discovery of G3425 provides evidence for the existence of mass-gap black holes in noninteracting binaries, which is hard to detect through X-ray emission.
Although the BSE simulations, using the stochastic prescription of the supernova-explosion mechanism, predict the existence of systems like G3425, the formation route of the wide circular orbit remains unclear.
Besides, G3425's selection from Gaia DR3 data, which observed 1.46 billion sources, covering only 1/100 of Galactic stars, suggests hundreds of such systems exist in the Milky Way.
This implies the BSE simulations may significantly underestimate the actual birth rate ($\approx$1 in the Galaxy) of G3425.
Future spectroscopic and astrometric observations, especially the upcoming Gaia DR4, may help unveil a low-mass black-hole binary population with a variety of parameters and provide profound insights into the formation and evolution of binary systems.

\begin{figure*}
\begin{center}
\includegraphics[width=1\textwidth]{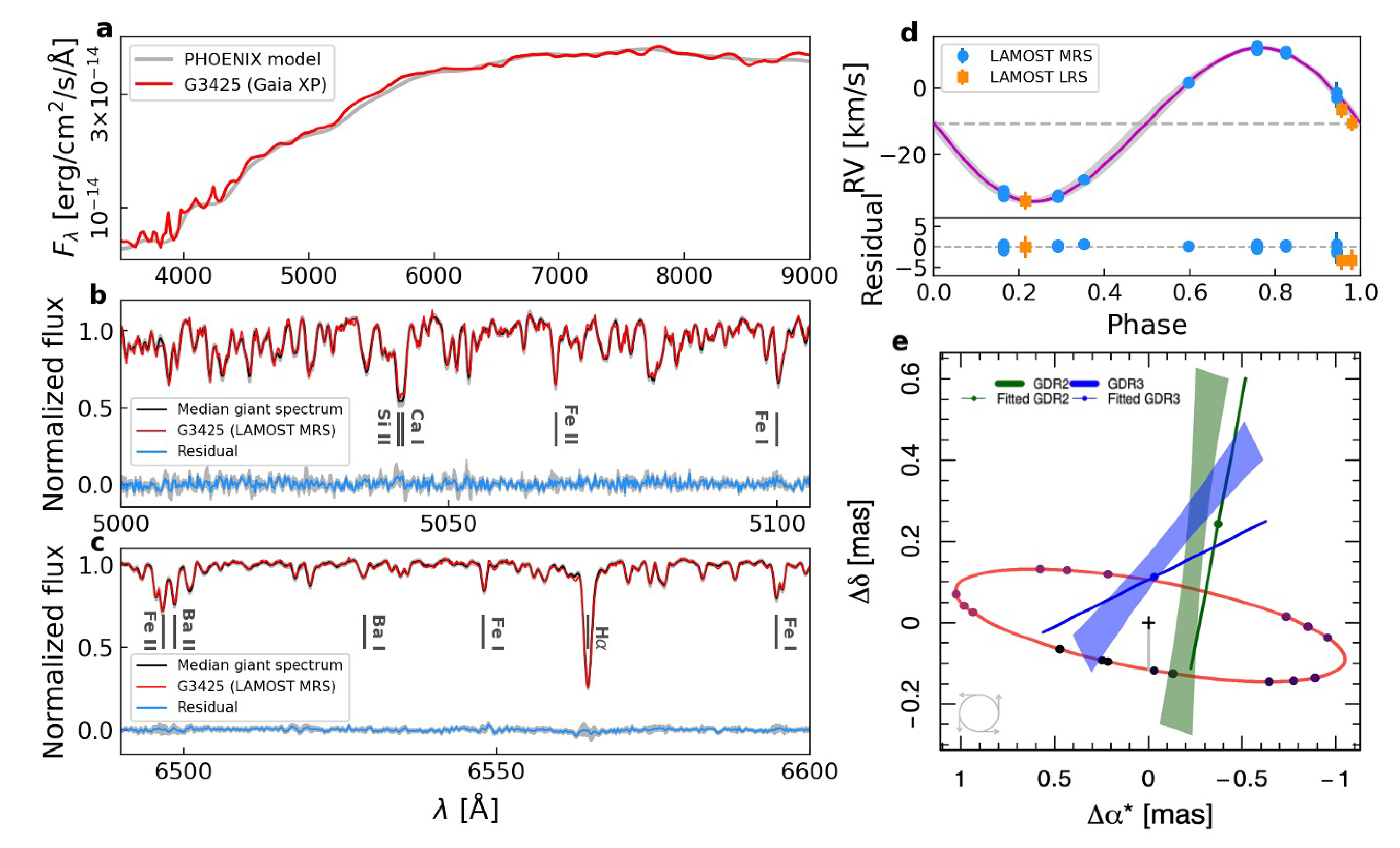}
\caption*{{\bf Fig. 1 $|$ Optical spectra and orbital motion fitting of the visible star.} (a) Re-calibrated Gaia XP spectrum (red; $R\approx30$) with one PHOENIX stellar template (grey; $T_{\rm eff} = $ 5000 K, log$g$=2.5, and [Fe/H] = 0) overplotted. (b) LAMOST blue-band medium-resolution spectrum (red; $R \approx 7500$) compared with a median spectrum (black) of about 800 giants with similar atomospheric parameters to G3425. The blue line indicates the residual. The shaded region represents 1$\sigma$ flux deviation among these giants' spectra.
(c) Same as panel (b) but for LAMOST red-band spectrum.
(d) Folded RV curve and binary orbital fits for the visible star. The RVs are obtained from LAMOST red-band low-resolution (orange; three points) and medium-resolution (blue; 27 points) spectra. The error bars represent 1$\sigma$ uncertainties.
The best-fit binary orbit model is marked with the purple line.
Shaded region represents the 1$\sigma$ confidence interval from MCMC results. Bottom panel shows the residual between the observed RV and the best-fit model. 
(e) Optimal fit to the Gaia Observing Schedule Tool (GOST) data, along with a comparison between best-fit and catalog proper motions and positions at Gaia DR2 (GDR2) and GDR3 reference epochs. The best-fit parameter values correspond to the maximum a posteriori (MAP) solution ($i\approx$ 84$^{\rm o}$) and may therefore slightly deviate from the median values reported in parameter tables (Methods). Shaded regions represent the uncertainty in Gaia position and proper motion. Each segment represents the best-fit position and proper motion offsets induced by the reflex motion at a specific reference epoch. The circle with arrows illustrates the direction of the Keplerian motion. The gradient of dot colors, transitioning from dark to bright, serves as a visual representation of the progression from early to late phases.} 
\label{motion.fig}
\end{center}
\end{figure*}

\begin{figure*}
\center
\includegraphics[width=1\textwidth]{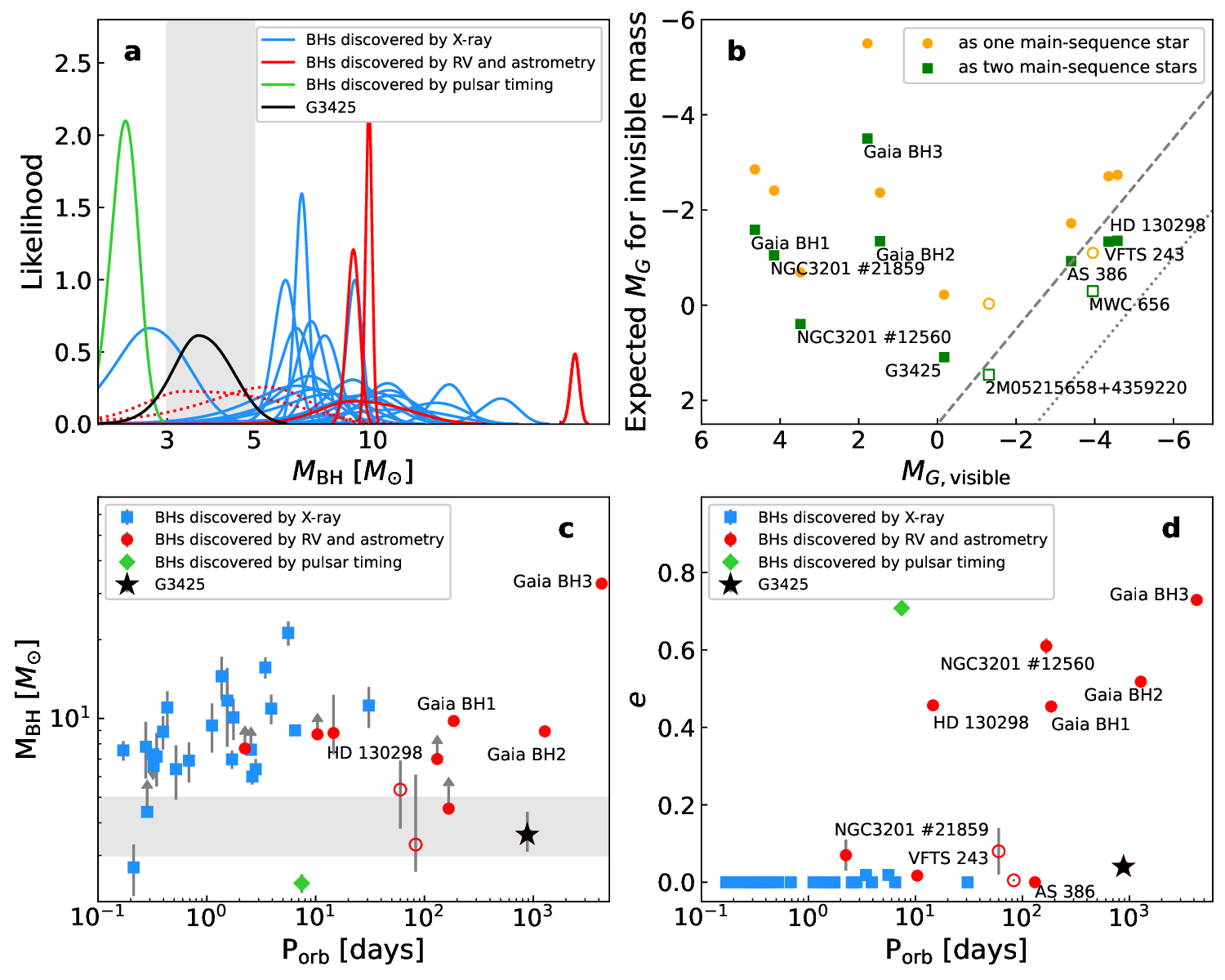}
\caption*{{\bf Fig. 2 $|$ Comparison of G3425 to other known black holes and candidates.}
Panel a: Likelihoods for the mass measurements of the black holes discovered by X-ray, RV and astrometry, and pulsar timing. Dotted red lines are for 2M05215658+4359220 and MWC 656, for which the status as black holes are uncertain. The shaded region marks the mass gap.
Panel b: Magnitudes of the visible star versus expected magnitudes for the invisible star as a main-sequence star (orange circles) or two main-sequence stars (green squares). Circles and squares with the same $M_{G, \rm visible}$ represent the same objects. The dashed and dotted lines represent a luminosity of 1/10 and 1/100 of the visible star, respectively. Objects below the lines may face difficulty distinguishing whether the invisible star is a black hole or not. Open symbols represent black hole candidates on debate, including MWC 656 and 2M05215658+4359220. 
Panel c: Orbital period versus black hole mass. Blue squares represent 24 black holes discovered by X-ray, while red circles represent 10 black holes and candidates discovered by RV and astrometry (Supplementary Information). The error bars represent 1$\sigma$ uncertainties. Objects with lower mass limits include GRS 1009-45, GS 1354-64, GC NGC3201 \#12560 and \#21859, VFTS 243, and AS 386.
Panel d: Orbital period versus eccentricity.}
\label{newbhs4.fig}
\end{figure*}

\clearpage

\begin{methods}

\subsection{Orbital fitting with RV data}

By using the zero-point corrected LAMOST RV data (Supplementary Information), we fit the Keplerian orbit of G3425 and obtain its orbital solution.
{\it The Joker}\cite{2017ApJ...837...20P}, a custom Markov chain Monte Carlo sampler, was employed for the fitting. 

First, we performed RV fitting in various period ranges, including 1--50 days, 50--200 days, 200--1200 days, and 1--1200 days.
No good fitting (i.e., the phase-folded RV points are not scattered) can be derived in the period ranges of 1--50 days or 50--200 days, while in the ranges of 200--1200 and 1--1200 days, the fittings converged to an orbital period of $\approx$880 days.
Next, we did a more precise fitting within the period range of 880$\pm$100 days.
Extended Data Fig. 1 displays the RV data along with the best-fit RV curves.
The difference between Gaia and our fitting is mainly caused by the short time line of Gaia observation, which only covers about one period. On the contrary, our observations span more than eight years, covering about four cycles, thereby enabling more accurate and reliable period estimation.
The scatter of the square points is due to the relatively low precision of RV measurements obtained from low-resolution data (also refer to the square points in Fig. 1d).
Extended Data Table 1 presents the orbital parameters of G3425 obtained through {\it The Joker} fitting, including orbital period $P$, eccentricity $e$, argument of periastron $\omega$, reference time $t_{\rm 0}$, RV semi-amplitude $K$, and systematic RV $\nu_{\rm 0}$.
Supplementary Fig. 1 shows the phased Mg I, Fe I, and H$_{\alpha}$ lines of the visible star.
It can be seen that G3425 displays an H$_{\alpha}$ absorption line at all orbital phases (Supplementary Fig. 2).

The binary mass function can be calculated as follows,
\begin{equation}
    f(M_2) = \frac{M_{2} \, \textrm{sin}^3 i} {(1+q)^{2}} = \frac{P \, K_{1}^{3} \, (1-e^2)^{3/2}}{2\pi G},
\end{equation}
where $M_{2}$ is the mass of the unseen star, $q = M_{1}/M_{2}$ is the mass ratio, and $i$ is the system inclination angle. 
For G3425, the mass function is $1.09\pm 0.02 M_{\odot}$.

\subsection{Combined analyses of RV and astrometry}
\label{astro.sec}

Since the parallax provided by Gaia DR3 is determined under the assumption of a single star, it may be unsuitable for wide binaries including compact components.
For example, the parallax values of Gaia BH2 differ between Gaia DR2, DR3, and the NTBO table, which are $1.58\pm0.03$, $0.67\pm0.10$, and $0.86\pm0.02$ mas, respectively.
Similarly, for G3425, Gaia DR2 reports a parallax value of $0.04\pm0.05$ mas while DR3 presents a value of $0.70\pm0.05$ mas.
To address the parallax inconsistency and to determine the inclination angle of the orbit, we analyzed the Gaia catalog data from both DR2 and DR3 as well as the RV data and sampled the posterior of the RV and astrometric models using the adaptive Markov Chain Monte Carlo (MCMC) adapted from the Delayed Rejection Adaptive Metropolis (DRAM) algorithm\cite{haario2006dram,2019ApJS..242...25F}.
This method has been successfully applied to constrain the orbits of the two nearest Jupiter-like planets, $\varepsilon$ Ind A b and $\varepsilon$ Eridani b\cite{2023MNRAS.525..607F}. 

Specifically, the RV model is a combination of a Keplerian component and the Moving Average (MA) model\cite{2013A&A...551A..79T} accounting for time-correlated noise. The logarithmic likelihood of the RV model is
\begin{equation}
 \ln{\mathcal{L}_{\rm rv}}=\sum_{i=1}^{N_{\rm rv}}\left[\ln{\frac{1}{\sqrt{2\pi(\sigma_i^2+J_{\rm rv}^2)}}}-\frac{(v_i-\hat{v}_i)^2}{2(\sigma_i^2+J_{\rm rv}^2)}\right]~,
\end{equation}
where $J_{\rm rv}$ is the excess RV noise, $\hat{v}_i\equiv v^{\rm kep}(t_i)+v^{\rm MA}(t_i)$ is the RV model and $v_i\equiv v(t_i)$ is the RV data at epoch $t_i$.
Simultaneously, by fitting a five-parameter model ($\hat{\vec\iota}$) to the synthetic abscissae of each Gaia DR through linear regression, the likelihood of the astrometry was calculated as follows:
\begin{align}
       \nonumber
    \ln{\mathcal{L}_{\rm gaia}}=&\ln\{(2\pi)^{-5/2}{\rm det}[\Sigma_{\rm GDR2}(1+J_{\rm gaia})]^{-\frac{1}{2}}\} \\\nonumber
    &-\frac{1}{2}(\hat\vec{\iota}_{\rm GDR2}-\vec{\iota}_{\rm GDR2})^T[\Sigma_{\rm GDR2}(1+J_{\rm gaia})]^{-1} \\\nonumber
    &(\hat\vec{\iota}_{\rm GDR2}-\vec{\iota}_{\rm GDR2}) \\\nonumber
    &+\ln\{(2\pi)^{-5/2}{\rm det}[\Sigma_{\rm GDR3}(1+J_{\rm gaia})]^{-\frac{1}{2}}\} \\\nonumber
    &-\frac{1}{2}(\hat\vec{\iota}_{\rm GDR3}-\vec{\iota}_{\rm GDR3})^T[\Sigma_{\rm GDR3}(1+J_{\rm gaia})]^{-1} \\
    &(\hat\vec{\iota}_{\rm GDR3}-\vec{\iota}_{\rm GDR3}) 
    ~,   
\end{align}  
where $\Sigma_{\rm GDR2}$ and $\Sigma_{\rm GDR3}$ are respectively the covariances of the GDR2 and GDR3 five-parameter astrometry, $J$ is the error inflation factor.
A detailed account of the model building is provided in Supplementary Information.

The total logarithmic likelihood of the combined model of RV and astrometry is 
\begin{equation}
    \ln\mathcal{L}=\ln{\mathcal{L}_{\rm rv}}+\ln{\mathcal{L}_{\rm gaia}}.
\end{equation}
Adopting log-uniform priors for time-related parameters, an informative prior for parallax, and uniform priors for other parameters, we launched multiple MCMC samplers to draw posterior samples\cite{haario2006dram}. 
We used two methods to derive the prior of parallax.
Firstly, we performed SED fitting with the PARSEC isochrones (\url{http://stev.oapd.inaf.it/cgi-bin/cmd\_3.1}).
We downloaded the sequences of isochrones with an age step of $\Delta$(log\,t) = 0.005, and collected all the points within the atmospheric parameter range ([$T_{\rm eff} - 100$ K, $T_{\rm eff} + 100$ K], [log$g$ $-$ 0.1, log$g$ $+$ 0.1]) as acceptable models.
For each model, we determined parallax and extinction values by fitting the observed magnitudes ($G$, BP, and RP bands of Gaia; $J$, $H$, and $K_{\rm S}$ bands of 2MASS; $B$ and $V$ bands of APASS).
The distribution of parallax results exhibited two peaks, with Gaussian fittings returning parallax values of 0.68$\pm$0.05 mas and 0.58$\pm$0.05 mas, respectively.
Secondly, we calculated spectroscopic masses using different parallaxes with a step of 0.01 mas, and compared them with the evolutionary mass estimated from the {\it isochrones} code\cite{2015ascl.soft03010M}.
A parallax of 0.62$\pm$0.05 mas/yr was considered a suitable prior, resulting in a spectroscopic mass equal to the evolutionary mass. 
Finally, a moderate prior of 0.6$\pm$0.1 mas/yr was applied by combining the three prior estimates.
Furthermore, we thoroughly explored the option of a priori correction for biases in Gaia DRs, such as parallax zero point and frame rotation. Our investigations revealed that these corrections did not result in any significant changes to the orbital solution.

The best fit of the RV and astrometry models to the data is shown in Extended Data Fig. 2. As shown in the second panel, the orbital phase is well sampled by Gaia observations. With GOST emulating Gaia epochs and refitting to the synthetic data\cite{2023MNRAS.525..607F}, we are able to constrain short period companions without approximating instantaneous position and velocity by catalog position and velocity. The decomposition of the best-fit astrometry model to the five-parameter solutions of Gaia DR2 and DR3 is shown in Extended Data Fig. 3. 
The 1D and 2D posterior distribution of orbital parameters are shown in Supplementary Fig. 3.
The model and data exhibit a close agreement within a 1-$\sigma$ confidence interval for nine astrometric parameters. 
The parameters for the combined RV and astrometry analyses are shown in Extended Data Table 2 and Supplementary Table 2. 
By subtracting the offset ($\Delta \varpi =$ 0.15 mas) from the Gaia DR3 parallax, we derived a new parallax of $0.56^{+0.09}_{-0.09}$ mas for G3425, corresponding to a distance of 1786$^{+342}_{-248}$ pc.
The fitted inclination angle was about $i = 89^{+15}_{-10}$ degrees, suggesting the orbit is close to edge-on. The uncertainty of the inclination angle does not significantly affect $M_2$ because $df(M_2)/di$ is approximately proportional to $M_2*{\rm sin}^2i*{\rm cos}i$.

In addition, as a test, we also applied these steps to Gaia BH2.
We derived the parallax priors of 0.81$\pm$0.1 mas/yr and 0.89$\pm$0.05 mas/yr, and a moderate parallax prior of 0.85$\pm$0.1 mas/yr.
We finally obtained a parallax of 0.84$^{+0.08}_{-0.05}$ mas, which is in agreement with the NTBO solution of $0.86\pm0.02$ mas.
The calculated inclination angle is $40.22_{-3.99}^{+7.43}$ degrees, consistent with previous estimation ($i$ = 34.9$\pm$0.4 degrees)\cite{2023MNRAS.521.4323E}.

\subsection{Mass of the visible star}

The stellar parameters of the visible star (Supplementary Information) are consistent with a giant star (Extended Data Fig. 4).
We used two methods to estimate its mass:

 (1) The {\it isochrones}\cite{2015ascl.soft03010M} module can be used to calculate the evolutionary mass by fitting the photometric or spectroscopic parameters with the Modules for Experiments in Stellar Astrophysics (MESA) models. 
We took the effective temperature $T_{\rm eff}$, surface gravity log$g$, metallicity [Fe/H], multi-band magnitudes ($G$, $G_{\rm BP}$, $G_{\rm RP}$, $J$, $H$, and $K_{\rm S}$), distance and extinction $A_V$ as the input parameters. 
The best-fit model yielded an evolutionary mass of $2.39^{+0.22}_{-0.18}\ M_{\odot}$ and a radius of $12.54_{-1.02}^{+0.85}\ R_{\odot}$.

 (2) We used six magnitudes ($G$, $G_{\rm BP}$, $G_{\rm RP}$, $J$, $H$, $K_{\rm S}$) and the atmospheric parameters to calculate the spectroscopic mass of G3425 following $M = gL_{\rm bol}/(4\pi G\sigma T_{\rm eff}^{4})$.
With the absolute luminosity and magnitude of the sun ($L_{\odot} =$ 3.83$\times$ 10$^{33}$ erg/s; $M_{\odot} =$ 4.74), the bolometric luminosity was calculated with
$L_{\rm bol} = 10^{0.4\times(M_{\odot} - M_{\rm bol})}L_{\odot}$.
%$M_{\odot} - M_{\rm bol} = 2.5\log(L_{\rm bol}/L_{\odot})$. 
%
The bolometric magnitude was calculated following
$m_{\rm \lambda} - M_{\rm bol} = 5{\rm log}d - 5 + A_{\rm \lambda} - BC_{\rm \lambda}$, where $m_{\rm \lambda}$ is the apparent magnitude of each band and $d$ is the distance from our new determination. 
$A_{\rm \lambda}$ is the extinction of each band calculated with $A_{\rm \lambda} = R_{\lambda} \times E(B-V)$, where $E(B-V) = 0.46\pm0.01$ was estimated from the StarPair method\cite{2013MNRAS.430.2188Y} and $R_{\lambda}$ is the extinction coefficient of each band\cite{1999PASP..111...63F}.
$BC_{\lambda}$ is the bolometric correction calculated using the {\it isochrones} package, with $T_{\rm eff}$, log$g$, and [Fe/H] values as the input.
The averaged spectroscopic mass from six bands is about $2.66^{+1.18}_{-0.68} M_{\odot}$, in agreement with the evolutionary mass estimation.

Note that different atmospheric parameters can lead to different mass estimates. Using the parameters from other methods (in Supplementary Table 3), we derived more spectroscopic mass estimates of the giant: $2.36^{+0.99}_{-0.62}$, $1.73^{+0.78}_{-0.49}$, $2.46^{+1.16}_{-0.73}$, $2.86^{+1.22}_{-0.76}$, and $2.31^{+0.97}_{-0.60}\ M_{\odot}$ with LASP/MRS, DD-Payne/LRS, SLAM/MRS, RRNet/MRS, and CYCLE-STAR/MRS, respectively.

Supplementary Table 4 lists the mass estimates of the giant star. 
Considering that the two stars may have interaction during the evolution of the binary system, especially during the early stage of progenitor of the black hole, the evolution of the visible star may deviate slightly from that of a single star. Therefore, the evolutionary mass estimated through isochrone fitting may be inaccurate to some extent. On the contrary, the spectroscopic mass is more reliable as it depends only on the star's current properties and not on its evolutionary history.

\subsection{Nature of the binary system}

The new derived parallax ($0.56^{+0.09}_{-0.09}$ mas/yr) corresponds to a distance of 1786$^{+342}_{-248}$ pc.
This leads to new spectroscopic and evolutionary mass measurements of $2.66^{+1.18}_{-0.68}$ and $2.39^{+0.22}_{-0.18}$ $M_{\odot}$ of the visible star, respectively.
The inclination angle was determined to be  $i = 89^{+15}_{-10}$ degrees, leading to the mass of the unseen object as $3.58^{+0.80}_{-0.47}$ $M_{\odot}$ or $3.42^{+0.16}_{-0.11}$ $M_{\odot}$ (Supplementary Table 5).
Additionally, different spectroscopic mass estimates of the giant from LASP/MRS, DD-Payne/
LRS, SLAM/MRS, RRNet/MRS, and CYCLE-STAR/MRS lead to different mass estimates of the unseen star, with values of $3.34^{+0.48}_{-0.54}$, $2.90^{+0.43}_{-0.48}$, $3.41^{+0.57}_{-0.64}$, $3.66^{+0.54}_{-0.65}$, and $3.31^{+0.49}_{-0.54} M_{\odot}$, respectively.

(1) {\it Whether the visible star is a stripped star with ongoing mass transfer?}
A prominent feature of binaries including stripped stars with ongoing mass transfer is that the filling factor of the stripped star is close to 1, such as V723 Mon\cite{2022MNRAS.512.5620E}, 2M0412\cite{2022MNRAS.516.5945J}, and NGC 1850 BH1\cite{2022MNRAS.511L..24E}.
The size of the Roche-lobe of G3425 can be calculated by using this approximation\cite{1983ApJ...268..368E}:
\begin{equation}
\frac{R_1}{a} =  
    \frac{0.49 q^{2/3}}{0.6q^{2/3} + \ln(1 + q^{1/3})},
\label{eq:qi} 
\end{equation} 
where $a = (1+q) a_1 = (1+q) P (1-e^2)^{1/2} K_1/(2 \pi {\rm \sin}i)$ is the separation of binary system. 
This approximation gives a Roche-lobe radius of $R_{\rm RL} = 290\pm29 R_\odot$
for G3425. According to its stellar radius ($\approx$13 $R_{\odot}$), the filling factor is only about $4.5\%$. 
This suggests that mass transfer through the Roche-lobe outflow is unlikely to occur.

G3425 displays an $H_{\alpha}$ absorption line at all orbital phases (Supplementary Fig. 2). However, it should be noted that binaries with ongoing mass transfer can display either emission lines (e.g., V723 Mon and 2M0412) or absorption lines (e.g., NGC 1850 BH1).
The $H_{\alpha}$ emission lines in V723 Mon and 2M0412 originate from possible accretion disks surrounding the hotter companions.
The $H_{\alpha}$ absorption line in NGC 1850 BH1 suggests a fast mass loss on a time-scale too short for the accretor to retain it\cite{2022MNRAS.511L..24E}.
Consequently, for NGC 1850 BH1, neither an accretion disk nor a decretion disk (around a star spun up to reach critical rotation) forms at such a low accretion rate; it is also possible that a very faint disk exists but is not detected.

(2) {\it Whether the visible star is a stripped star with post mass transfer?}
HR 6819 and LB-1 have been debated to be binaries including a stripped star that underwent post mass transfer\cite{2021MNRAS.502.3436E,2020A&A...639L...6S}. Both of them include a hot subgiant donor star and a Be accretor. They exhibit $H_{\alpha}$ emission lines originating from a decretion disk around the Be star. Below, we present evidence that G3425 did not follow the stripped-star evolutionary path like these binaries.

First, a stripped star after the termination of mass transfer would enter into a contraction phase with nearly constant luminosity. At such a phase, the effective temperature increases a lot.
For example, the temperature would rise about 2--3 times when the star’s radius shrinks to less than 0.1 of its Roche lobe. However, in the case of G3425, the ratio of the star's radius to the Roche lobe is about 0.045, indicating that the post-mass-transfer stripped star scenario does not apply to G3425.

Second, assuming the star has recently undergone envelope stripping, which means it had overflowed its Roche lobe with an orbit close to the current one, it must have had a radius $R \gtrsim 200 R_{\odot}$.
This excludes a low-mass ($M \lesssim 0.5 M_{\odot}$) stripped giant, because a giant of this mass would not have reached such a large radius to overflow its Roche lobe in the G3425 system\cite{1995MNRAS.273..731R}.
For a stripped star with a mass of 0.5/1.0/1.5/2.0 $M_{\odot}$, the inferred mass of the companion is about 1.79/2.26/2.66/3.02 $M_{\odot}$.
For a main-sequence star with above masses, the $G$-band magnitude is about 2.16/0.93/0.62/0.29 mag, corresponding a luminosity ratio of 0.12/0.37/0.49/0.67 compared to the visible star.
This means the main-sequence star can be detected in the SED or spectrum.
However, our analysis reveals no excess in the blue band of the SED and spectra.

Third, to search for possible anomalous spectral features (caused by mass transfer), we compared the spectrum of G3425 with giants having similar stellar parameters and abundances observed by LAMOST.
The selection criteria included: 4700 $K$ $<T_{\rm eff}<$ 5000 $K$; 2.4 $<{\rm log}g<$ 2.6; -0.5 $<$[Fe/H]$<$ 0; S/N $>$ 50; the number of observations $N_{\rm obs}\geq3$ and $\Delta RV<5$ km/s; Galactocentric Cartesian  coordinates of $-15$ kpc $<X<-$5 kpc, $-$5 kpc $<Y<5$ kpc, and $-$2 kpc $<Z<$ 2 kpc. 
The Galactocentric Cartesian coordinate of G3425 is ($X$, $Y$, $Z$) $=$ ($-$9.89, $-$0.25, 0.14) kpc.
We derived 30,333 spectra from 794 giants observed with LAMOST MRS.
For each spectrum, we preformed resampling of the wavelength from 5000 \AA\ to 5100 \AA and from 6500 \AA\ to 6600 \AA, both with 2000 points.
We obtained a median spectrum by connecting the median value of each wavelength point.
Fig. 1 shows a comparison between the median giant spectrum and the spectrum of G3425 with highest S/N.
Our analysis reveals quite good agreement between the two observed spectra and reveals no excess absorption or emission in spectral lines (e.g.,  H$_{\alpha}$), excluding the light contribution of an additional component.
We also measured the abundances of 15 elements (Supplementary Table 3), which show no anomalies for a red giant star.
In addition, the lithium (Li) abundance, defined as
$A$(Li) $=$ log(N$_{\rm Li}/$N$_{\rm H}$) $+$ 12, is only about 1.1\cite{2021ApJ...914..116G}, indicating  there is no Li enhancement in G3425.

Forth, we used the Binary Population and Spectral Synthesis code (BPASS) to search for post-interaction binaries exhibiting properties similar to G3425.
The search criteria included:
(i) Age = [0.1, 1] Gyr; (ii) $P = 880 \pm 200$ day; (iii) $J = -1.486 \pm 1$ mag; (iv) $H = -2.017 \pm 1$ mag; (v) $K = -2.083 \pm 1$ mag; (vi) $V = 0.011 \pm 1$ mag; (vii) log($T_{{\rm 1}}$) = 3.5--3.8 $K$; (viii) $M_{{\rm 1}}$ = 0-5 $M_{\odot}$; (ix) $M_{{\rm 2}}$ = 0--5 $M_{\odot}$.
We first examined the selected models including compact objects, referred to as ``secondary models". All these models consist of a normal giant with a mass of 2--3 $M_{\odot}$ and a compact object (i.e., a white dwarf or neutron star). These models don't involve a stripped star, although the progenitor of the compact object may have been stripped. In these models, the giant is considerably more massive than the compact object, in contrast to the binary mass function of the G3425 system.
Then, we investigated the selected models with two normal stars, referred to as ``primary models". We found none of these binaries had entered into the mass transfer stage when they met above criteria.

(3) {\it Whether the unseen object is a main-sequence star?}
For the visible star, the absolute magnitude in the $G$ band is about -0.19 mag (with distance $= 1786$ pc); while for a main-sequence star with a mass of 3.6 $M_{\odot}$, its $G$-band absolute magnitude is about $-$0.2 mag (\url{http://www.pas.rochester.edu/~emamajek/EEM\_dwarf\_UBVIJHK\_colors\_Teff.txt}).
The comparable luminosity of those two components suggests that there must be a flux excess in the blue band if the component star is a normal star.
Extended Data Fig. 5 shows the comparison of the observed spectrum---the flux-calibrated Gaia XP spectrum, and the PHOENIX\cite{Husser2013} template spectrum. 
There is no flux excess in blue band, indicating the companion star is a compact object if it has a mass of 3.6 $M_{\odot}$.

(4) {\it Whether the unseen object is a binary including two equal-mass main-sequence stars?}
In the case of a binary consisting of two main-sequence stars with a mass of $\approx$1.8 $M_{\odot}$, the total $G$-band absolute magnitude is around 1.35 mag. 
The $G$-band luminosity ratio for the close binary and the visible giant of G3425 can be calculated to be approximately 1/4.
Furthermore, the $BP$-band absolute magnitude for the close binary is around 1.4 mag, and the corresponding luminosity ratio is about 1/3.
Such a binary would also exhibit a noticeable signal in the blue band of the spectra.

(5) {\it Whether the binary spectrum includes two components?}
Spectral disentangling is the most effective method to identify the secondary star in a binary system when it has a comparable luminosity to the primary star.
In previous studies, many black hole candidates were identified as binaries consisting of a stripped donor star and a main-sequence accretor, based on the spectral disentangling method.

In this work, we used a linear algebra technique\cite{1994A&A...281..286S} to disentangle spectra.
Before applying this algorithm, we performed preliminary tests to determine the detection limits for G3425. 
Assuming the secondary as a main-sequence star with different masses ($\approx$3.9, 3.6, 3.4, 2.7, 2.3, 1.9 $M_{\odot}$ corresponding a serious of mass ratios of $q=$ 0.7, 0.75, 0.8, 1, 1.2, 1.4), we derived its effective temperature and surface gravity of the secondary (\url{http://www.pas.rochester.edu/~emamajek/EEM\_dwarf\_UBVIJHK\_colors\_Teff.txt}).
The BT-COND model\cite{2012RSPTA.370.2765A} with the closest atmosphere parameters is used as the theoretical spectrum of the secondary star.
The luminosity ratios of the giant to the secondary star were 0.33, 0.44, 0.61, 1.44, 2.28, 7.04, respectively.
We then combined the observed LAMOST MRS spectra of G3425 and the theoretical spectrum with a resolution of $R=$ 7500 and various rotational broadening ($V{\rm sin}i =$ 10 km/s, 50 km/s, 100 km/s and 150 km/s) to get synthetic binary spectra.
The spectral disentangling was performed in the wavelength range of 5000 to 5200 \AA\  (blue band) and 6400 to 6600 \AA\ (red band) using a luminosity ratio of 0.48 (i.e., the flux ratio in $G$ band).
Supplementary Table 6 shows that the spectra in the blue band can be successfully separated when $V{\rm sin}i$ is lower than 150 km/s, while the spectra in the red band can be well separated even when $V{\rm sin}i$ is higher than 150 km/s.

We carried out spectral disentangling on the observed LAMOST MRS of G3425 using the same wavelength ranges as previous tests. 
The mass ratios used were 0.7, 0.75, 0.8, 1, 1.2, 1.4, respectively.
Extended Data Fig. 6 displays one example (with $q=$ 0.75) of the results of spectral disentangling for three different epochs.
In all these steps, no additional component with an absorption feature can be observed as from the other visible component (the red lines in Extended Data Fig. 6), indicating that the secondary of G3425 is a dark compact object.

We also used the Fourier domain-based disentangling code FDBinary (new version: fd3)\cite{2004ASPC..318..111I}, with the help of a PYTHON wrapper (\url{https://github.com/ayushmoharana/fd3\_initiator}), to perform spectral disentangling.
The used spectral wavelength ranges from 5000 to 5200 ${\rm \AA}$.
We tried two modes by treating G3425 as a single-lined binary (i.e., component A is present while component B is absent) or double-lined binary (i.e., both component A and B are present), respectively (\url{http://sail.zpf.fer.hr/fdbinary/}). 
If G3425 is a normal binary, the residual of the single-lined binary mode would show absorption features, while the double-lined binary mode would produce two stellar spectra.
The orbital period and eccentricity were set to be 881$\pm$10 days and 0.04$\pm$0.02, respectively.
The semi-amplitude of RV of star A was set to be 23.3$\pm$1 km/s.
In the case of a double-lined binary, the RV semi-amplitude of star B was calculated using the mass ratio ($\approx$0.75) and set to be 17.5$\pm$1 km/s.
In both modes, no clear signal from any other component was detected, further indicating the companion is a compact object.

\subsection{Formation scenario}

The evolutionary scenario of G3425 needs to explain the low mass of the black hole and the wide circular orbit. 

(1) Isolate binary evolution scenario. Assuming the black hole in this binary system formed from an individual star, the mass of the black hole depends on three major factors: initial stellar mass, mass loss/transfer during the star’s life, and the final core-collapse/supernova process. 
The mass loss/transfer and collapse/explosion processes are also responsible for the fate of the binary orbit.
The dynamical capture scenario can be first rejected, since the tidal circulation timescale is much longer than the Hubble time. Given the black hole's low mass, we conducted a series of evolutionary models with the masses of the initial progenitor stars set around 15 $M_{\odot}$, which represents the lower limit for black hole formation\cite{2016ApJ...821...38S}. The helium cores formed by these models have a mass of about 5--7 $M_{\odot}$, indicating these stars need to lose almost their entire hydrogen envelope before the collapse stage. 
Considering the high mass ratio between the black hole progenitor and the visible giant in G3425, the mass transfer is likely unstable, leading to the common envelope phase. We ran a series of simulations using an adiabatic mass-loss model\cite{2010ApJ...717..724G,2015ApJ...812...40G,2020ApJ...899..132G} for the binary evolution. To successfully eject the common envelope, our simulations showed that the binding energy needs to reduce to 1/10 of the original value, or the external energy input needs to be 10 times the orbital energy (Supplementary Information). 

(2) Triple system scenario. Triple systems are common in the universe, particularly for high-mass early O-type or B-type primaries, where the fraction of triples can reach several tens of percent\cite{2017ApJS..230...15M}. In this scenario, the progenitor of G3425 includes the observed giant star as an outermost component and an inner binary containing two massive stars.
For the inner binary, the more massive one evolves first, forming a neutron star.
The system would evolve into a common envelope stage and the neutron star sinks towards the star's core before the common envelope is ejected.
The inner binary possibly forms a massive Thorne-\.{Z}ytkow object\cite{1995MNRAS.274..485P}.
The present black hole finally formed when a significant fraction of the envelope was accreted onto the core. 
Alternatively, it is possible that the common envelope was successfully ejected before the merge, and the central unseen object is the merger product of two neutron stars or even still contains two neutron stars (Supplementary Information).

\subsection{Binary population synthesis}

We used a population synthesis code BSE\cite{2002MNRAS.329..897H} to simulate the formation of binary systems similar to G3425. This code has been modified, including the treatments of stellar winds, mass transfer processes, common envelope evolution, and supernova explosion mechanisms\cite{2014ApJ...796...37S}. If G3425 is indeed formed via a binary channel, its progenitor system must have experienced a common envelope phase. During this phase, the orbital energy of the binary is used to eject the shared envelope originating from the donor star. In our simulations, we employed two different ejection efficiencies $\alpha_{\rm CE}$ for common envelope evolution, i.e., $\alpha_{\rm CE} = 1$ and $\alpha_{\rm CE} = 5$. Considering that the black hole of G3425 has the mass in the mass gap, we adopt two supernova-explosion mechanisms involving the delayed\cite{2012ApJ...749...91F} and the stochastic\cite{2020MNRAS.499.3214M} prescriptions to deal with the compact remnant masses and the natal kicks. Both the two mechanisms have been incorporated in the BSE code\cite{2021ApJ...920...81S}. 

Each simulation contains $\sim$ 10$^7$ binary systems that begin the evolution from the primordial binaries with two zero-age main sequence stars. For the parameter configurations of primordial binaries, we assumed that the masses of the primary stars obey the Kroupa's initial mass function\cite{1993MNRAS.262..545K}, the mass ratios of the secondary to the primary stars follow a flat distribution between 0 and 1\cite{2007ApJ...670..747K}, the orbital separations are uniformly distributed in the logarithmic space\cite{1983ARA&A..21..343A}, and the eccentricities have a thermal-equilibrium distribution between 0 and 1\cite{1991A&A...248..485D}. In our simulations, we took the primary masses in the range of 5--100 $M_{\odot}$, the secondary masses in the range of 0.3--100 $M_{\odot}$, and the orbital separations in the range of 3--10000 $R_{\odot}$. For the stars formed in the Milky Way, we adopt a constant star formation rate of 3\,$M_{\odot}\,{\rm yr}^{-1} $ with solar metallicity over a period of 10 Gyr.

We identified the binary systems with properties similar to G3425 with following conditions: (1) $500 \,{\rm days} < P_{\rm orb} < 1500 \,{\rm days}$; (2) $3.1\ M_{\odot} < M_{\rm BH} < 4.1\ M_{\odot}$; (3) $1.7\ M_{\odot} < M_{\rm giant} < 3.7\ M_{\odot}$; (4) $0 < e < 0.1$. These binaries are detached black hole systems with a (sub)giant companion of mass $M_{\rm giant}$. There are two big differences between our adopted supernova-explosion mechanisms of forming black holes. On the one hand, the minimum mass of black hole’s progenitors can reach as low as $\sim 12M_{\odot}$ in the stochastic prescription\cite{2021ApJ...920...81S}, which is smaller than that in the delayed prescription.  On the other hand, the stochastic prescription involves relatively small natal kicks with a fraction receiving zero kicks, while the delayed prescription involves significantly large natal kicks. Based on our simulations, we estimated that the total number of Galactic detached black hole binaries with a (sub)giant companion decreases from $\sim 2000$ in the stochastic prescription to $\sim 100$ in the delayed prescription, which is not sensitive to the options of common envelope ejection efficiencies between $\alpha_{\rm CE} = 1$ and $\alpha_{\rm CE} = 5$. 
Extended Data Fig. 7 shows the calculated number distributions of the binary systems with $3.1\ M_\odot < M_{\rm BH} < 4.1\ M_\odot$ and $1.7\ M_\odot < M_{\rm giant} < 3.7\ M_\odot$ in the plane of the orbital period versus eccentricity. 
There are four panels corresponding to different assumptions for supernova-explosion mechanisms and common envelope ejection efficiencies. In each panel, the black star marks the potion of G3425. We found that binaries similar to G3425 can only be formed if adopting the stochastic prescription of the supernova-explosion mechanism. Even in this mechanism, we expected that G3425-like binaries are very rare systems in the Milky Way. Their number increases from $\sim$0.02 to $\sim$1 when varying common envelope ejection efficiencies from $\alpha_{\rm CE} = 1$ to $\alpha_{\rm CE} = 5$. After analyzing our calculated data, we obtained that the corresponding primordial binaries of G3425-like systems have the primary masses of $\sim$13--20 $M_{\odot}$, the secondary masses of $\sim$1.5--3.5 $M_{\odot}$, and the orbital periods of $\sim$2000--10000 days. 

\clearpage

\begin{table}
\caption*{\bf Extended Data Table 1 $|$ Kepler orbital parameters from {\it The Joker} fitting to LAMOST RV data and from the Gaia NTBO catalog. \label{keplerorbit.tab}}
\centering
\setlength{\tabcolsep}{3mm}
 \begin{tabular}{cccc}
\hline\noalign{\smallskip}
Parameters & LAMOST RV$_{\rm b,cor}$ & LAMOST RV$_{\rm r,cor}$ & Gaia DR3\\
\hline\noalign{\smallskip}
$P$ (day) & $881.22^{+1.79}_{-1.81}$ & $876.53^{+2.27}_{-2.19}$ & $914.84\pm42.44$ \\
$e$ & $0.04^{+0.01}_{-0.01}$ & $0.05^{+0.01}_{-0.01}$ & $0.13\pm0.09$ \\ 
$\omega$ (rad) & $1.63^{+0.20}_{-0.23}$ & $1.78^{+0.15}_{-0.17}$ & $2.30\pm0.21$ \\ 
%$t{\rm 0}$ & $2457407.12193$ & $2457407.12193$ & $2457514.38752\pm27.02336$ \\ 
$t{\rm 0}$ (BJD-2457400) & $7.12^{+28.05}_{-32.25}$ & $7.12^{+26.70}_{-30.92}$ & $114.39\pm27.02$ \\ 
$K_{\rm 1}$ (km/s) & $23.28^{+0.10}_{-0.09}$ & $22.89^{+0.11}_{-0.11}$ & $22.79\pm0.76$ \\ 
$\nu{\rm 0}$ (km/s) & $-10.86^{+0.12}_{-0.12}$ & $-10.66^{+0.16}_{-0.16}$ & $-10.73\pm0.70$ \\ 
%\hline
$f(m)\ (M_{\odot})$ & $1.15^{+0.02}_{-0.02}$ & $1.09^{+0.02}_{-0.02}$ & $1.08^{+0.13}_{-0.12}$ \\ 
%$M_{\rm 2,min}^{\rm evo}\ (M_{\odot})$ & $3.38^{+0.13}_{-0.14}$ & $3.28^{+0.13}_{-0.14}$ & $3.25^{+0.25}_{-0.24}$ \\ 
%$M_{\rm 2,min}^{\rm gra}\ (M_{\odot})$ & $3.57^{+0.55}_{-0.64}$ & $3.46^{+0.55}_{-0.63}$ & $3.44^{+0.59}_{-0.64}$ \\ 
\noalign{\smallskip}\hline
\end{tabular}
\end{table}

\begin{table}
\caption*{\bf Extended Data Table 2 $|$ Parameters for G3425 from the joint fitting to LAMOST RV and Gaia DR2 and DR3 data.}
\label{tab:solution}
\begin{center}
\small
\renewcommand{\arraystretch}{0.85}
\begin{tabular}{llp{3.5cm}llp{1.3cm}p{1.4cm}}
\hline
Parameter$^a$ &Unit & Meaning& Value$^c$ & Prior$^d$ & $\theta_{\rm min}$ ($\mu$) & $\theta_{\rm max}$ ($\sigma$)\\
\hline
$P$&day&Orbital period&$879.11_{-2.57}^{+3.22}$ & Log-Uniform&-1&16\\
$K_{\rm 1}$&km/s&RV semi-amplitude&$22.91_{-0.14}^{+0.13}$ & Uniform&$10^{-6}$&$10^{6}$\\
$e$&---&Eccentricity&$0.05_{-0.01}^{+0.01}$ & Uniform&0&1\\
$i$&deg&Inclination&$89.30_{-10.08}^{+15.48}$ & CosI-Uniform&-1&1\\
$\omega$&deg&Argument of periastron$^b$&$105.70_{-12.66}^{+10.24}$ & Uniform&0&2$\pi$\\
$\Omega$&deg&Longitude of ascending node&$271.95_{-26.38}^{+27.22}$ & Uniform&0&2$\pi$\\
$M_0$&deg&Mean anomaly at JD 2457407.12193&$337.24_{-10.69}^{+14.58}$ & Uniform&0&2$\pi$\\\hline
$\Delta \alpha*$&mas&$\alpha*$ offset&$-0.14_{-0.14}^{+0.13}$ & Uniform&$-10^6$&$10^6$\\
$\Delta \delta$&mas&$\delta$ offset&$0.22_{-0.13}^{+0.12}$ & Uniform&$-10^6$&$10^6$\\
$\Delta \varpi$&mas&$\varpi$ offset&$0.15_{-0.09}^{+0.09}$ & Gaussian&$0.1$&$0.1$\\
$\Delta \mu_{\alpha*}$&mas/yr&$\mu_{\alpha*}$ offset &$0.41_{-0.24}^{+0.22}$ & Uniform&$-10^6$&$10^6$\\
$\Delta \mu_\delta$&mas/yr&$\mu_\delta$ offset&$-0.23_{-0.21}^{+0.23}$ & Uniform&$-10^6$&$10^6$\\
\hline
$P$&yr&Orbital period &$2.41_{-0.01}^{+0.01}$ & ---&---&---\\
$a$&au&Semi-major axis &$3.31_{-0.21}^{+0.32}$ & ---&---&---\\
$M_{\rm 2}$&$M_\odot$&Companion mass &$3.58_{-0.47}^{+0.80}$ & ---&---&---\\ %$m_c$
$T_p-2400000$&JD&Periastron epoch&$57462.81_{-35.72}^{+26.02}$ & ---&---&---\\
$M_{\rm 1}$&$M_\odot$&Mass of the primary&$2.66_{-0.68}^{+1.18}$ & ---&---&---\\
$\varpi$&mas&Parallax&$0.56_{-0.09}^{+0.09}$ & ---&---&---\\
\hline
\multicolumn{7}{p{1.0\textwidth}}{\footnotesize$^a$ The first 12 rows
  show parameters that are inferred directly through MCMC posterior
  sampling, while the last five rows show the parameters derived from
  the directly sampled parameters. The mass of the visible star $M_{\rm 1}$ is derived from the parallax posterior and the spectroscopic properties. The semi-major axis $a$ and companion mass $m_c$ are derived from the orbital parameters and $M_{\rm 1}$. The parallax $\varpi$ is determined by $\varpi_{\rm GDR3}-\Delta \varpi$. }\\
    \multicolumn{7}{p{1.0\textwidth}}{\footnotesize$^b$ This is the argument of periastron of the reflex motion of the visible star and $\omega+\pi$ is the argument of periastron of companion orbit.}\\
    \multicolumn{7}{p{1.0\textwidth}}{\footnotesize$^c$ The optimal value of a parameter aligns with the median of its posterior distribution, while the uncertainty is represented by the 1$\sigma$ confidence interval. }\\
  \multicolumn{7}{p{1.0\textwidth}}{\footnotesize$^d$ The rightest three
  columns show the prior distribution and the corresponding minimum (or mean) and maximum (or standard deviation) for a parameter. ``Log-Uniform'' is the logarithmic uniform
  distribution, and ``CosI-Uniform'' is the uniform distribution over
  $\cos I$.}\\
 \end{tabular}
\end{center}
\end{table}

\clearpage

\begin{figure}
    \center
    \includegraphics[width=1\textwidth]{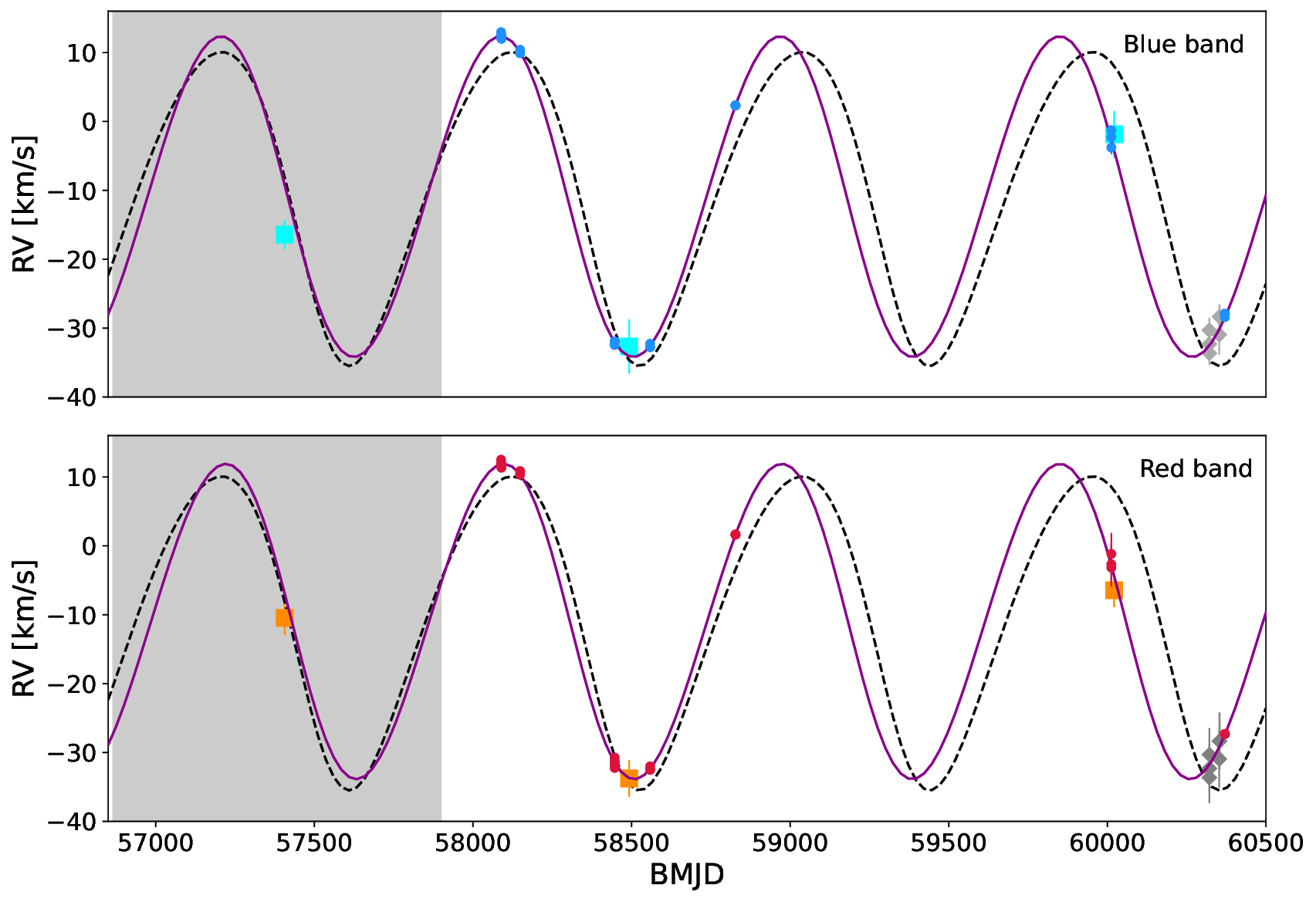}
    \caption*{{\bf Extended Data Fig. 1 $|$ Best-fit RV curves.} Top panel: best-fit RV curve (purple line) using the blue-band RV data, compared with the RV curve from Gaia NTBO catalog (dashed line). The RV data from LAMOST LRS and MRS and 2.16 m telescope are marked by three cyan squares, 27 blue dots, and five grey diamonds, respectively. The error bars represent 1$\sigma$ uncertainties. The RV data given by 2.16 m telescope on BJD$\approx$2460322 were combined to enhance the accuracy.
    The grey area shows the observation time of the data used by Gaia, from July 25th 2014 to May 28th 2017. Bottom panel: best-fit RV curve (purple line) using the red-band RV data, compared with the RV curve from Gaia NTBO catalog (dashed line). The RV data from LAMOST LRS and MRS and 2.16 m telescope are marked by orange squares, red dots, and grey diamonds, respectively.}
    \label{blue_red_joker.fig}
\end{figure}

\begin{figure}
    \center   
    \includegraphics[width=1\textwidth]{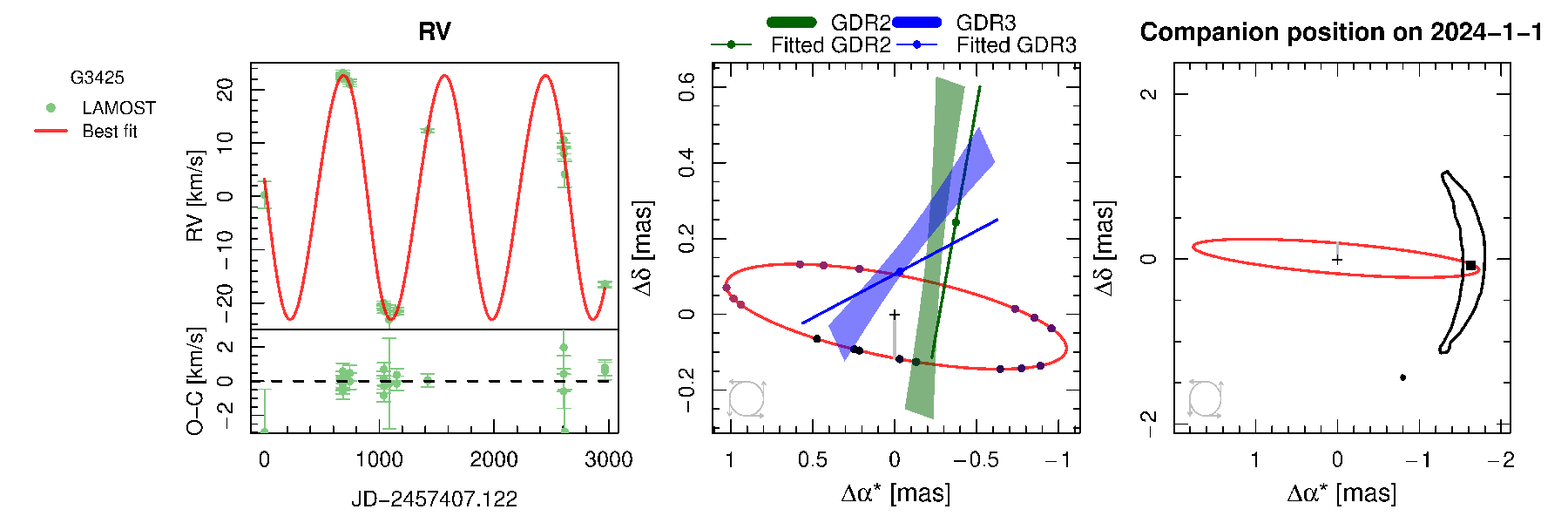}
    \caption*{{\bf Extended Data Fig. 2 $|$ Optimal orbital solutions for G3425.} The panels from left to right respectively display the best fits to LAMOST RV data, Gaia GOST data, and the predicted G3425 position on January 1st, 2024. In the left panel, the best-fit RV curve is represented by the red line, overplotted with the LAMOST LRS and MRS RV data (30 green points). The error bars represent 1$\sigma$ uncertainties. In the middle panel, the optimal fit to the Gaia GOST data is shown, along with a comparison between best-fit and catalog proper motions and positions at Gaia DR2 (GDR2) and GDR3 reference epochs. The shaded regions represent the uncertainty in position and proper motion. Each segment represents the best-fit position and proper motion offsets induced by the reflex motion at a specific reference epoch. The right panel displays the predicted companion position on January 1st, 2024, with a 1-$\sigma$ contour line to indicate the prediction uncertainty.}
    \label{fig:fit1}
\end{figure}

\begin{figure}
    \center   
    \includegraphics[width=1\textwidth]{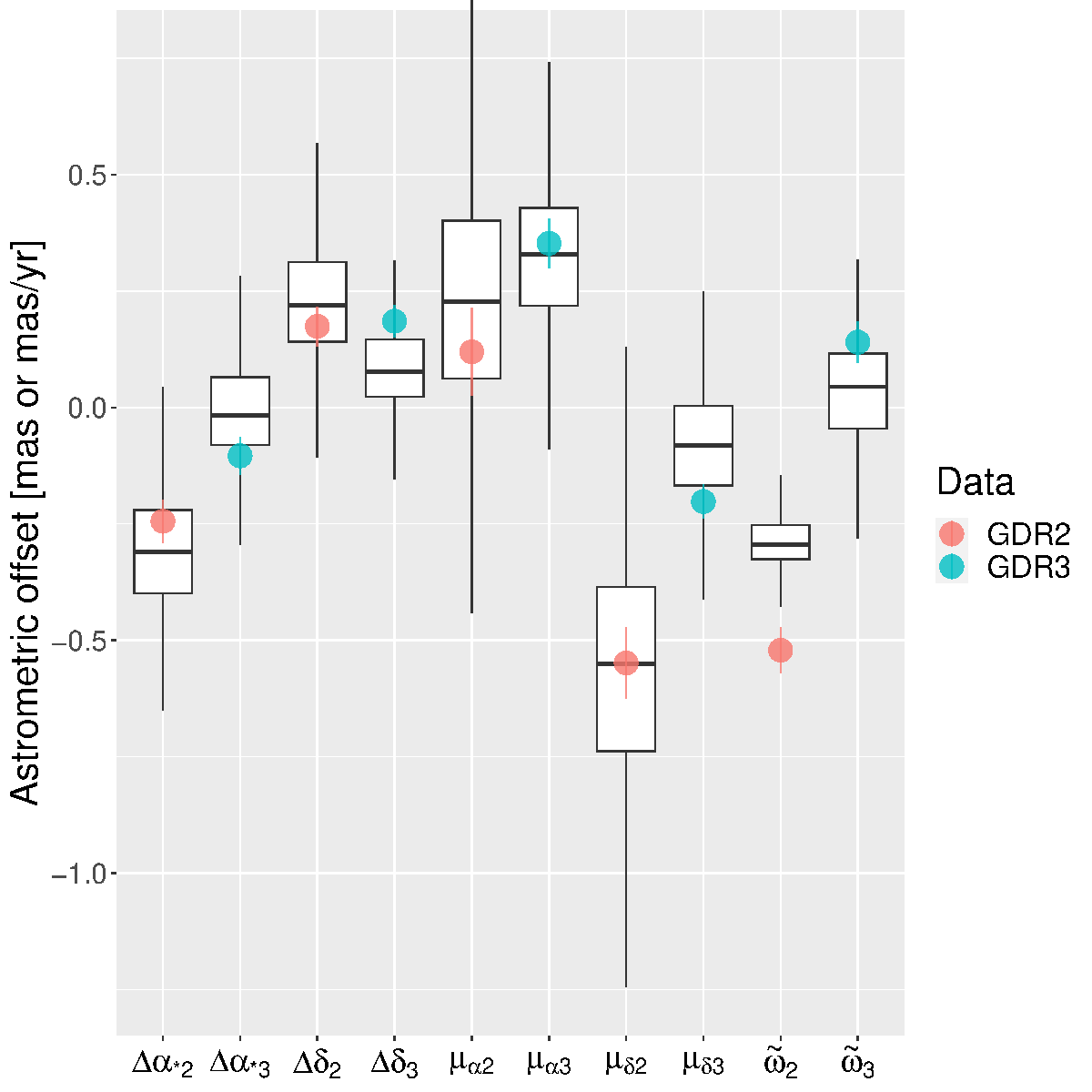}
    \caption*{{\bf Extended Data Fig. 3 $|$ Model fit to the five-parameter astrometry of GDR2 and GDR3.} The astrometry of the barycenter is subtracted from the five-parameter solutions (represented by dots with error bars) and from the model predictions based on posterior sampling (represented by box plots). For the box plot, the lower and upper hinges correspond to the 25th and 75th percentiles. The upper whisker extends from the hinge to the largest value within 1.5*IQR (inter-quartile range). The lower whisker extends to the smallest value within 1.5*IQR. Data beyond the whiskers are ``outliers".}
    \label{fig:fit2}
\end{figure}

\begin{figure}
    \center
    \includegraphics[width=1\textwidth]{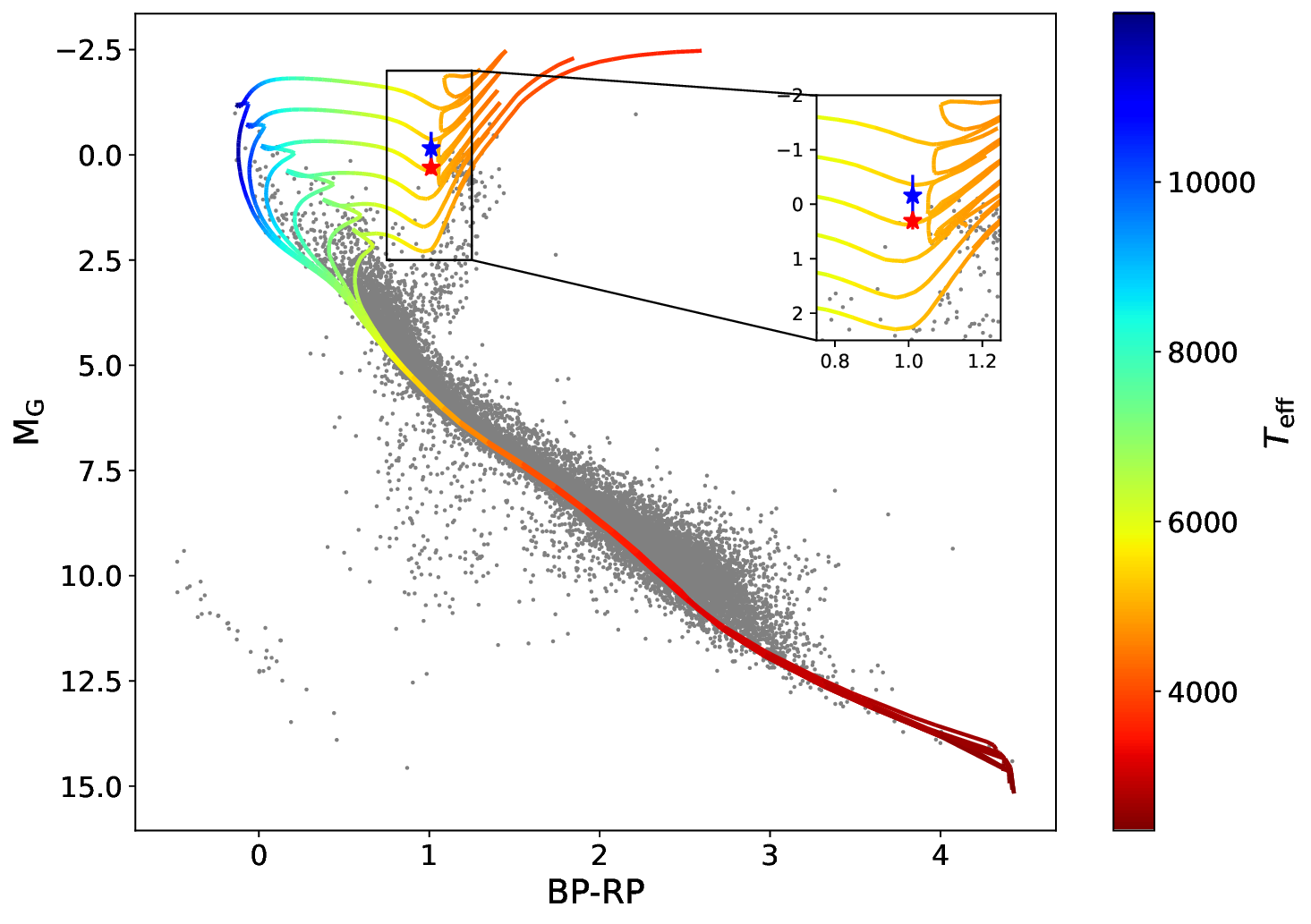}
    \caption*{{\bf Extended Data Fig. 4 $|$ Position of G3425 in Hertzsprung–Russell diagram.} The isochrones (from top to bottom) from PARSEC models have an age of $10^{8.4}$/$10^{8.6}$/$10^{8.8}$/$10^{9.0}$/$10^{9.2}$/$10^{9.4}$ yr, respectively. The red and blue stars are calculated with the distances from Gaia DR3 (1442 pc) and our joint fitting (1786 pc), respectively. The grey background points are selected from Gaia DR2 with distances of $d <$ 200 pc, $G_{\rm mag}$ between 5--16 mag, and galactic latitude $|b|$ $>$ 10. No extinction correction was applied to these background stars.}
    \label{HR_diagram.fig}
\end{figure}

\begin{figure}
    \center
    \includegraphics[width=1\textwidth]{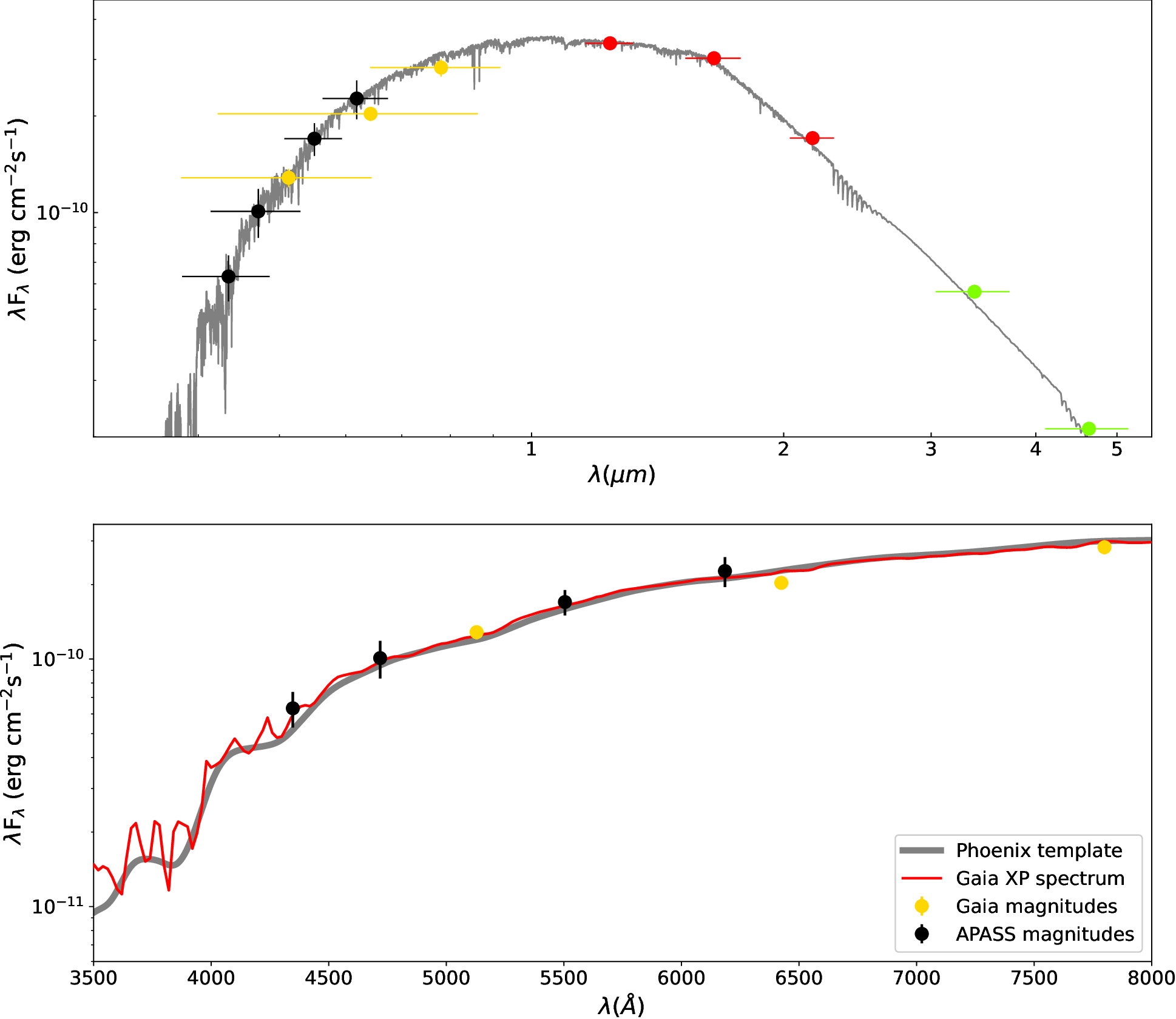}
    \caption*{{\bf Extended Data Fig. 5 $|$ SED and Gaia XP spectrum of G3425.}
    Top panel: SED fitting for G3425 with a distance of 1786$^{+342}_{-248}$ pc.
    The observed data come from APASS (three black dots), Gaia (three yellow dots), 2MASS (three red dots) and WISE (two green dots). The grey line is the best-fit model. The error bars represent 1$\sigma$ uncertainties. 
    Bottom panel: Comparison of the observed spectrum (flux-calibrated Gaia XP spectrum) and the PHOENIX template spectrum ($T_{\rm eff} = 5000$ K, log$g = 2.5$, [Fe/H]=0.0), with multi-band magnitudes overplotted.
    }
    \label{single_sed.fig}
\end{figure}

\begin{figure*}
    \center
    \includegraphics[width=1\textwidth]{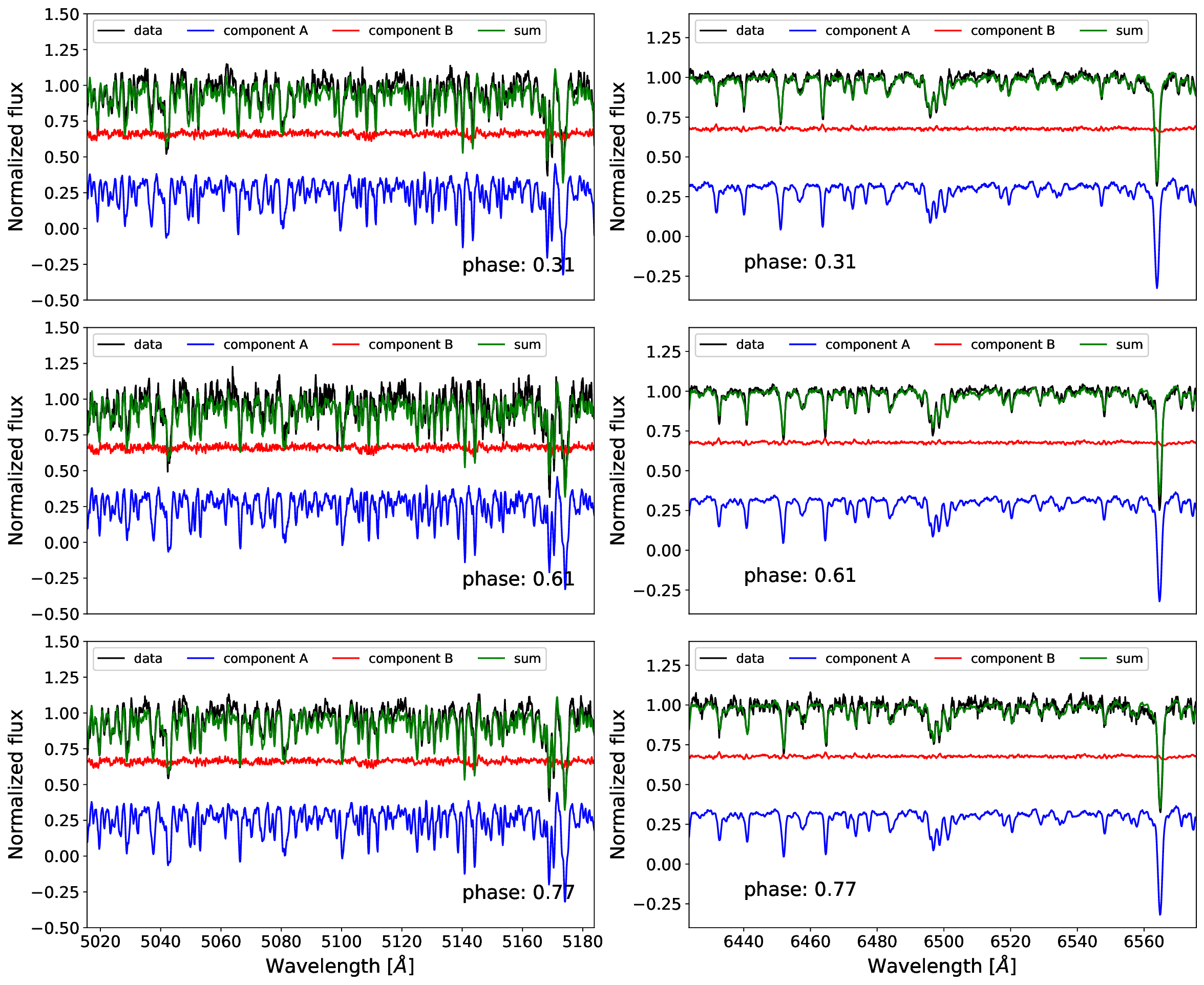}
    \caption*{{\bf Extended Data Fig. 6 $|$ Example of spectral disentangling ($q =$ 0.75) of G3425 using the blue band and red band of the LAMOST MRS.} The vertical panels show the spectra at phases of 0.31, 0.61, and 0.77. The blue lines are the separated spectra of the visible star (primary); the red lines show the spectra from the other component; the green lines represent the sum of blue lines and red lines. The black lines are the spectra from LAMOST MRS observations.}
    \label{dis_3425.fig}
\end{figure*}

\begin{figure*}
    \center
    \includegraphics[width=1.\textwidth]{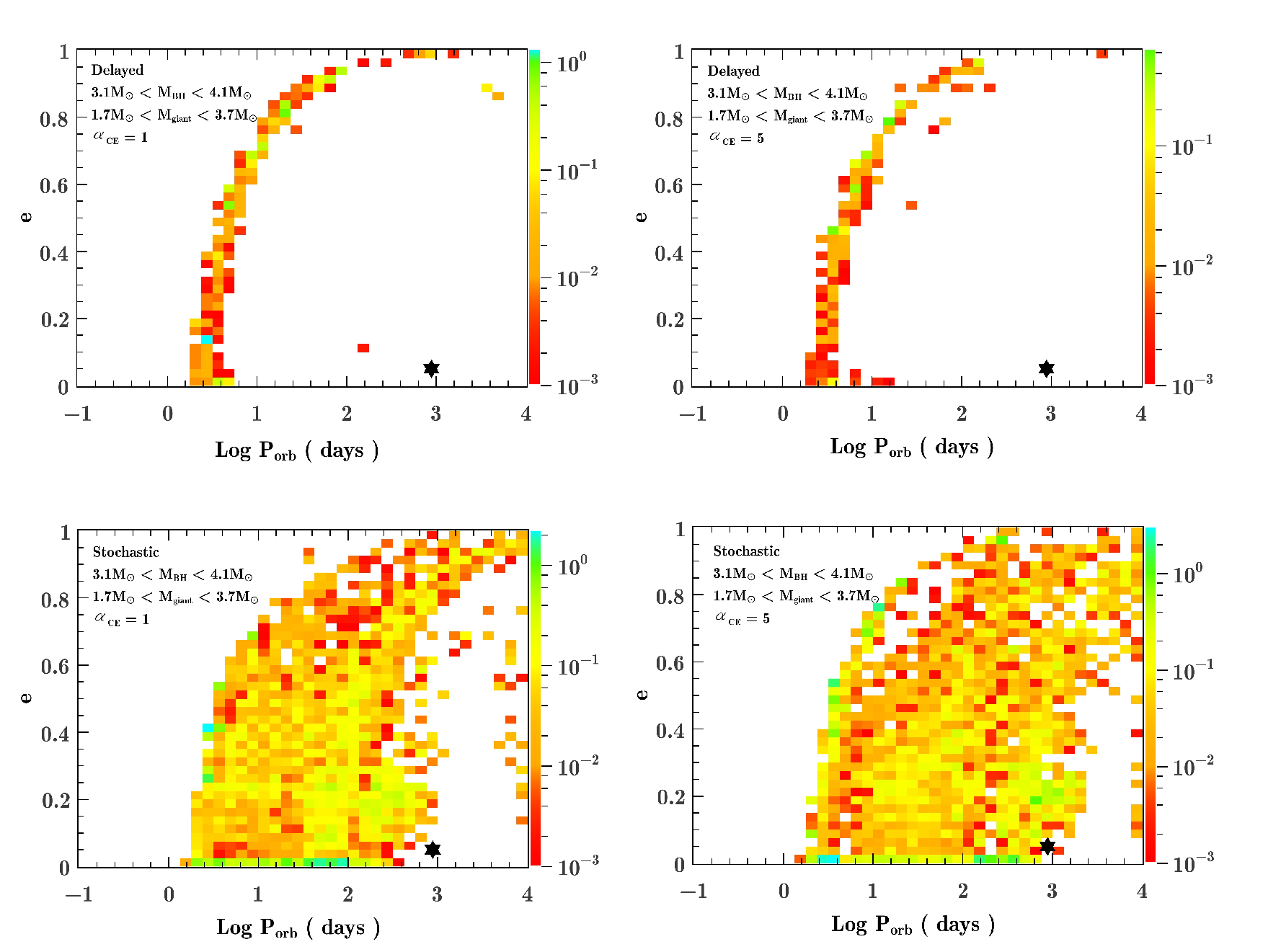}
    \caption*{{\bf Extended Data Fig. 7 $|$ Estimated number distribution of the Galactic detached black hole binaries with a (sub)giant companion in the plane of orbital period versus eccentricity.} Only the systems with $3.1\ M_\odot < M_{\rm BH} < 4.1\ M_\odot$ and $1.7\ M_\odot < M_{\rm giant} < 3.7\ M_\odot$ are included. The top and bottom panels correspond to the delayed and stochastic supernova-explosion mechanisms, respectively. The left and right panels correspond to common envelope ejection efficiencies of $\alpha_{\rm CE} = 1$ and $\alpha_{\rm CE} = 5$, respectively. In each panel, the black star marks the position of G3425.}
   \label{BHs.fig}
\end{figure*}

\end{methods}

\clearpage

\begin{addendum}

\item[Data availability]

The LAMOST spectra used in this paper are available from LAMOST database: \url{https://www.lamost.org/lmusers}. The RV data, stellar parameters of the visible star, and parameters of known black holes are listed in tables in Supplementary Information. The other data that support the plots within this paper and other findings of this study are available from the corresponding authors upon reasonable request.

\item[Acknowledgements]

This work is supported by the National Science Foundation of China (NSFC) under grant numbers 11988101/11933004 (J.F.L), 12273057 (S.W.), 12041301/12121003 (X.D.L.), U2031205/12233002 (Q.Z.L.), 12288102/12125303 (X.F.C.), 12173081 (H.W.G.), 123B2045 (S.J.G.), 12041303 (P.W.), 12333008 (X.C.M.), 12473066 (F.B.F.), and 12103086 (Zhenwei Li)
It is also supported by the National Key Research and Development Program of China (NKRDPC) under grant numbers 2023YFA1607900 (W.M.G.), 2023YFA1607901 (S.W.), 2023YFA1607902 (Y.S.), 2016YFA0400804 (J.F.L.), 2021YFA0718500 (X.D.L.), 2021YFA0718500 (Q.Z.L.), 2021YFA1600403 (X.F.C.).
It is also supported by Strategic Priority Program of the Chinese Academy of Sciences under grant No. XDB41000000 (S.W.), XDB0550300 (Y.S.). 
F.B.F. thanks the Shanghai Jiao Tong University 2030 Initiative.
H.W.G. thanks the key research program of frontier sciences, CAS, No. ZDBS-LY-7005, Yunnan Fundamental Research Projects (grant No. 202101AV070001).
X.C.M. thanks the Yunnan Fundamental Research Projects (grant nos. 202401BC070007 and 202201BC070003) and the International Centre of Supernovae, Yunnan Key Laboratory (grant no. 202302AN360001). Zhenwei Li thanks the Yunnan Fundamental Research Projects (grant no. 202401AT070139). P.W. thanks the CAS Youth Interdisciplinary Team, the Youth Innovation Promotion Association CAS (id. 2021055), and the Cultivation Project for FAST Scientific Payoff and Research Achievement of CAMS-CAS. 
This work is made possible with LAMOST (Large Sky Area Multi-Object Fiber Spectroscopic Telescope), a National Major Scientific Project built by the Chinese Academy of Sciences. Funding for the project has been provided by the National Development and Reform Commission. LAMOST is operated and managed by the National Astronomical Observatories, Chinese Academy of Sciences. 
This work presents results from the European Space Agency (ESA) space mission Gaia. Gaia data are being processed by the Gaia Data Processing and Analysis Consortium (DPAC). Funding for the DPAC is provided by national institutions, in particular the institutions participating in the Gaia MultiLateral Agreement (MLA). 
We acknowledge the support of the staff of the Xinglong 2.16m telescope. This work was partially Supported by the Open Project Program of the CAS Key Laboratory of Optical Astronomy, National Astronomical Observatories, Chinese Academy of Sciences.

\item[Author Contributions]

S.W. and X.L.Z. contributed equally to this work.
S.W., F.B.F. and J.F.L. are equally responsible for supervising the discovery.
S.W. selected G3425 from Gaia DR3 sample.
S.W. and X.L.Z. reduced the LAMOST data, performed RV and stellar parameter analysis. 
F.B.F. did the joint fitting of RV and astrometric data.
H.W.G., Y.S., Y.Z.C., S.J.G., and L.F.Z. performed binary evolution simulations.
S.W. and X.L.Z. wrote the manuscript with help mainly from J.F.L., F.B.F., H.W.G., Y.S., Y.Z.C., S.J.G., and Y.H.
P.W., X.L., Z.R.B., H.L.Y., H.B.Y., Z.X.Z., T.Y., J.B.Z., T.D.L., M.S.X., H.G.H., M.Z. and D.W.F. also contributed to data analysis.
X.D.L., X.F.C., Zhengwei Liu, X.C.M., Q.Z.L., H.T.Z., W.M.G., and Zhenwei Li also contributed to the binary formation interpretation and discussion. 
All contributed to the paper in various forms.

\item[Competing interests]

The authors declare no competing interests.

\item[Additional information] 

{\bf Correspondence and requests for materials} should be addressed to S. Wang (email: songw@bao.ac.cn), F.B. Feng (email: ffeng@sjtu.edu.cn) and/or J.F. Liu  (email: jfliu@nao.cas.cn).

{\bf Peer review information} Nature Astronomy thanks the anonymous reviewers for their contribution to the peer review of this work.

{\bf Reprints and permissions information} is available at www.nature.com/reprints.

\end{addendum}

\clearpage
%%%%%%%%%%%%%%%%%%%%%%%%%%%%%%%%%%%%%%%%%%%%%%%%%%
%references 2; references for Methods
%%%%%%%%%%%%%%%%%%%%%%%%%%%%%%%%%%%%%%%%%%%%%%%%%%

%\bibliography{bibliography.bib}
%\printbibliography

%\includepdf[pages=-]{Wang_SI.pdf}

\clearpage

\section*{\large Supplementary Information}

\subsection{Discovery}\label{intro.sec}

We selected compact object candidates from the Gaia DR3 {\sc nss\_two\_body\_orbit} (hereafter NTBO) catalog\cite{2023A&A...674A..34G}, which includes 443,205 binary systems identified from spectroscopy, astrometry, or photometry. 
For many binaries (e.g., ``AstroSpectroSB1", ``SB1", or ``SB2"), the NTBO table provides orbital solutions including period $P$, eccentricity $e$, semi-amplitude $K$ of the primary star, and possibly the inclination angle $i$, etc. 

First, we selected binaries flagged as ``AstroSpectroSB1", ``SB1", ``SB1c", or ``Orbital", and cross-matched them with the LAMOST DR9 low-resolution and medium-resolution general catalogue (\url{http://www.lamost.org/dr9/v1.0/catalogue}).
We then picked out sources with two more spectroscopic observations, each with a signal-to-noise ratio (SNR) higher than 5, and clear RV variation.
Second, we calculated the binary mass function of those systems using orbital parameters from the Gaia NTBO table, and determined the gravitational masses of visible stars using atmospheric parameters from LAMOST. 
Third, we estimated the minimum masses of the unseen stars, and the sources with a minimum mass higher than $\approx$1 $M_{\odot}$ would be studied in detail.
A group of sources have been picked out and studied\cite{2024ApJ...964..101Z}.
For Gaia DR3 3425577610762832384, the NTBO table presents an orbital solution including an orbital period $P$ = 914$\pm$42 days, a center-of-mass velocity $V_{\rm 0}$ = $-$10.73$\pm$0.70 km s$^{-1}$, a semi-amplitude $K_{\rm 1}$ $=$ 22.79$\pm$0.76 km s$^{-1}$, and an eccentricity $e$ $=$ 0.13$\pm$0.09.
An estimate of the binary mass function from Gaia solution is about 1.08$\pm$0.12.

\subsection{RV measurements and fitting}

We calculated the RV of each spectrum using the cross-correlation function (CCF).
For MRS, we used the entire blue and red bands, excluding the initial and final 400 data points for each band, to calculate the RVs (i.e., RV$_{\rm b}$ and RV$_{\rm r}$), respectively.
For LRS, we utilized the blue band spectrum ranging from 4500 \AA\ to 6000 \AA\ and the red band spectrum ranging from 6300 \AA\ to 8000 \AA. 
The PHOENIX model with similar atmospheric parameters was used as the template.
Due to the temporal variation of the zero-points, small systemic offsets exist in RV measurements\cite{2019RAA....19...75L,2020ApJS..251...15Z}.
Therefore, the RV value of each spectrum (i.e., each fiber at each exposure) needs a correction with corresponding zero point.
Here, we used the Gaia DR3 data to determine the RV zero points (RVZPs) for each spectrograph night by night, and applied them as the common RV shift of the fibers in the same spectrograph.
For each spectrograph, we selected common sources in Gaia DR3 with some criteria ({\it ruwe} $<$ 1.4; {\it radial\_velocity} !$=$ 0; {\it rv\_amplitude\_robust} $<$ 10 km/s) to exclude possible binaries, and then compared the RVs of the common objects in each exposure and those from Gaia DR3, and determined a median offset $\Delta RV$ for one night with two or three iterations\cite{2021RAA....21..292W}. 
Finally, we took the square root of the sum of the squares of the measurement error, the wavelength calibration error, and the error of RVZP as the RV’s uncertainty.
Supplementary Table 1 listed the results of RV measurements from low-resolution spectra (LRS) and medium-resolution spectra (MRS), including RV$_{\rm b}$ and RV$_{\rm r}$ being RVs from CCF measurement and RV$_{\rm b,cor}$ and RV$_{\rm r,cor}$ being RVs with RVZP correction.

Besides, we performed five observations using the Beijing Faint Object Spectrograph and Camera (BFOSC) mounted on the 2.16 m telescope at the Xinglong Observatory. 
The observed spectra were reduced using the IRAF v2.16 software\cite{1986SPIE..627..733T} following standard steps, and the reduced spectra were then corrected to vacuum wavelength.

The NTBO table contains orbital information of G3425, including period $P$, eccentricity $e$, semi-
amplitude $K$ of the primary star, etc (Extended Data Table 1).
Here we used both the RV$_{\rm b,cor}$ and RV$_{\rm r,cor}$, estimated from LAMOST low- and medium-resolution observations, to fit the Keplerian orbit of G3425 and obtain its orbital solution.
{\it The Joker}\cite{2017ApJ...837...20P}, a custom Markov chain Monte Carlo sampler, was employed for the fitting. 

First, we performed RV fitting in various period ranges, including 1--50 days, 50--200 days, 200--1200 days, and 1--1200 days.
No good fitting (i.e., the phase-folded RV points are not scattered) can be derived in the period ranges of 1--50 days or 50--200 days, while in the ranges of 200--1200 and 1--1200 days, the fittings converged to an orbital period of $\approx$880 days.
Next, we did a more precise fitting within the period range of 880$\pm$100 days.
Extended Data Fig. 1 displays the RV data along with the best-fit RV curves.
The difference between Gaia and our fitting is mainly caused by the short time line of Gaia observation, which only covers about one period. On the contrary, our observations span more than eight years, covering about four cycles, thereby enabling more accurate and reliable period estimation.
The scatter of the square points is due to the relatively low precision of RV measurements obtained from low-resolution data (also refer to the square points in Fig. 1d).
Extended Data Table 1 presents the orbital parameters of G3425 obtained through {\it The Joker} fitting, including orbital period $P$, eccentricity $e$, argument of periastron $\omega$, reference time $t_{\rm 0}$, RV semi-amplitude $K$, and systematic RV $\nu_{\rm 0}$.
Supplementary Fig. 1 shows the phased Mg I, Fe I, and H$_{\alpha}$ lines of the visible star.
It can be seen that G3425 displays an H$_{\alpha}$ absorption line at all orbital phases (Supplementary Fig. 2).

The binary mass function can be calculated as follows,
\begin{equation}
    f(M_2) = \frac{M_{2} \, \textrm{sin}^3 i} {(1+q)^{2}} = \frac{P \, K_{1}^{3} \, (1-e^2)^{3/2}}{2\pi G},
\end{equation}
where $M_{2}$ is the mass of the unseen star, $q = M_{1}/M_{2}$ is the mass ratio, and $i$ is the system inclination angle. 
For G3425, the mass function is $1.09\pm 0.02 M_{\odot}$.

\subsection{Combined analyses of RV and astrometry}
\label{astro.sec}

Since the parallax provided by Gaia DR3 is determined under the assumption of a single star, it may be unsuitable for wide binaries including compact components.
For example, the parallax values of Gaia BH2 differ between Gaia DR2, DR3, and the NTBO table, which are $1.58\pm0.03$, $0.67\pm0.10$, and $0.86\pm0.02$ mas, respectively.
Similarly, for G3425, Gaia DR2 reports a parallax value of $0.04\pm0.05$ mas while DR3 presents a value of $0.70\pm0.05$ mas.
To address the parallax inconsistency and to determine the inclination angle of the orbit, we analyzed the Gaia catalog data from both DR2 and DR3 as well as the RV data and sampled the posterior of the RV and astrometric models using the adaptive Markov Chain Monte Carlo (MCMC) adapted from the DRAM algorithm\cite{haario2006dram,2019ApJS..242...25F}.
This method has been successfully applied to constrain the orbits of the two nearest Jupiter-like planets, $\varepsilon$ Ind A b and $\varepsilon$ Eridani b\cite{2023MNRAS.525..607F}. 

Specifically, the RV model is a combination of a Keplerian component and the Moving Average (MA) model\cite{2013A&A...551A..79T} accounting for time-correlated noise. The Keplerian RV for the $k^{\rm th}$ companion is 
%\begin{equation}
\begin{align}
\nonumber
  v^{\rm kep}_k(t_i)=&\sqrt{\frac{Gm_{ck}^2}{(m_{ck}+m_s)a_k(1-e_k^2)}}\\
  &\sin{I_k}\left[\cos{(\omega_k+\nu_k(t_i))}+e_k\cos(\omega_k)\right]+b~,
     \end{align}
     %\label{eqn:rv2}
%\end{equation}
where $b$ is the RV offset and independent offsets are adopted for different instruments, $m_{ck}$ is the mass, $a_k$, $e_k$, $I_k$, $\omega_k$ and $\nu_k$ are respectively the semi-major axis, eccentricity, inclination, argument of periastron, and true anomaly of the stellar reflex motion induced by the $k^{\rm th}$ companion. The total Keplerian RV is 
\begin{equation}
    v^{\rm kep}(t_i)=\sum_{k=1}^{N} v^{\rm kep}_k(t_i).
\end{equation}
The MA model is
\begin{equation}
  v^{\rm MA}(t_i)=\sum_{l=1}^q w_l {\rm
    exp}\left(-\frac{|t_i-t_{i-l}|}{\tau}\right)\left[v(t_{i-l})-v^{\rm kep}(t_{i-l})\right]~,
  \label{eqn:rv1}
\end{equation}
where $v(t_{i-l})$ is the RV data at time $t_{i-l}$, $\tau$ is the amplitude and timescale of the $q^{\rm th}$ order MA model (or MA(q)), $w_l$ is the amplitude of the $l^{\rm th}$ component of MA(q). 

To avoid over-fitting, it is important to choose the optimal noise model to account for time-
correlated noise in RV data\cite{2016MNRAS.461.2440F}. We chose the best MA model by finding the highest order $q$ that increases the logarithmic Bayes factor by at least 5 in the Bayesian framework\cite{2017MNRAS.470.4794F}. The MA(1) model is found to be optimal for noise modeling. The logarithmic likelihood of the RV model is
\begin{equation}
 \ln{\mathcal{L}_{\rm rv}}=\sum_{i=1}^{N_{\rm rv}}\left[\ln{\frac{1}{\sqrt{2\pi(\sigma_i^2+J_{\rm rv}^2)}}}-\frac{(v_i-\hat{v}_i)^2}{2(\sigma_i^2+J_{\rm rv}^2)}\right]~,
\end{equation}
where $J_{\rm rv}$ is the excess RV noise, $\hat{v}_i\equiv v^{\rm kep}(t_i)+v^{\rm MA}(t_i)$ is the RV model and $v_i\equiv v(t_i)$ is the RV data at epoch $t_i$.

For unresolved binaries, Gaia measures the motion of the system photocenter around the mass center
rather than the reflex motion of the star. For a mass-luminosity function of $F(m)$, the angular size of the
photocenter motion is
\begin{equation}
a_p=\frac{a}{d}\frac{m_2}{m_1+m_2}\left[1-\frac{m_1F_2(m_2)}{m_2F_1(m_1)}\right],
\end{equation}
where $m_1$ and $m_2$ are, respectively, the masses of the primary and the secondary, $a$ is the semi-major axis, $d$ is the distance which
is derived from parallax by $d=A_u/\varpi$ where $A_u\equiv1$\,au. Because the companion studied in this work does not contribute significant flux to the photocenter, we assumed $F(m_2)=0$. Hence the motion of the photocenter is equivalent to the stellar reflex motion and has an amplitude of $a_r=a_p=\frac{a}{d}\frac{m_2}{m_1+m_2}$. 

The stellar reflex motion induced by a companion in the orbital plane is 
\begin{eqnarray}
  x_i&=&\cos{E_i}-e~,\nonumber\\
  y_i&=&\sqrt{1-e^2}\sin{E_i}~,\\
  \label{eq:orbital}
\end{eqnarray}
where $E_i$ is the eccentric anomaly at time $t_i$, $e$ is the eccentricity. The reflex motion in the sky plane is 
\begin{eqnarray}
  \Delta\alpha_{*i}^r&=&BX_{i}+GY_{i}~,\\
  \Delta\delta_i^r&=&AX_{i}+FY_{i}~,
\end{eqnarray}
where $A$, $B$, $F$, $G$ are the Thiele-Innes elements\cite{1883AN....104..245T}, and are functions of $a_p$, inclination $I$, argument of periastron of the star $\omega_*$, and longitude of ascending node $\Omega$. 

The astrometric parameters of a single star are right ascension (R.A.; $\alpha$), declination (decl.; $\delta$), parallax ($\varpi$), proper motion in R.A. and in decl. ($\mu_\alpha$ and $\mu_\delta$). We defined $\alpha_*=\alpha\cos\delta$ and $\vec{\iota}=(\alpha_*,\delta,\varpi,\mu_\alpha,\mu_\delta)$. The Gaia DR2 and DR3 catalog data are modeled through the following steps:
  \begin{itemize}
      \item we corrected DR3 astrometry to barycentric astrometry by subtracting astrometric offsets, $\vec{\iota}_{\rm bary}=\vec{\iota}_{\rm GDR3}-\Delta\vec{\iota}$~;
      \item we obtained the scan angles and along-scan parallax factors at the Gaia observational epochs generated by the Gaia Observation Forecast Tool (GOST; \url{https://gaia.esac.esa.int/gost/}) ; 
      \item we transformed the barycentric astrometry from the celestial coordinate system to the Cartesian coordinate system and then propagate the barycentric astrometry from DR3 epoch to GOST epochs linearly by assuming a constant barycentric motion; 
      \item we generated the synthetic Gaia abscissa at each GOST epoch as follows:
      \begin{eqnarray}
          \eta_i=(\alpha_{*i}+\Delta\alpha_{*i}^r)\sin\psi_i+(\delta_i+\Delta\delta_i^r)\cos\psi_i+\varpi_i f_i^{\rm AL}, 
      \end{eqnarray}
      where $f_i^{\rm AL}$ is the parallax factor at epoch $t_i$, $\psi_i$ is the scan angle;
      \item we fit a five-parameter model ($\hat{\vec\iota}$) to the synthetic abscissae of each Gaia DR through linear regression and calculate the likelihood as follows:
       \begin{align}
       \nonumber
    \ln{\mathcal{L}_{\rm gaia}}=&\ln\{(2\pi)^{-5/2}{\rm det}[\Sigma_{\rm GDR2}(1+J_{\rm gaia})]^{-\frac{1}{2}}\} \\\nonumber
    &-\frac{1}{2}(\hat\vec{\iota}_{\rm GDR2}-\vec{\iota}_{\rm GDR2})^T[\Sigma_{\rm GDR2}(1+J_{\rm gaia})]^{-1} \\\nonumber
    &(\hat\vec{\iota}_{\rm GDR2}-\vec{\iota}_{\rm GDR2}) \\\nonumber
    &+\ln\{(2\pi)^{-5/2}{\rm det}[\Sigma_{\rm GDR3}(1+J_{\rm gaia})]^{-\frac{1}{2}}\} \\\nonumber
    &-\frac{1}{2}(\hat\vec{\iota}_{\rm GDR3}-\vec{\iota}_{\rm GDR3})^T[\Sigma_{\rm GDR3}(1+J_{\rm gaia})]^{-1} \\
    &(\hat\vec{\iota}_{\rm GDR3}-\vec{\iota}_{\rm GDR3}) 
    ~,   
    \end{align}  
    where $\Sigma_{\rm GDR2}$ and $\Sigma_{\rm GDR3}$ are respectively the covariances of the GDR2 and GDR3 five-parameter astrometry, $J$ is the error inflation factor.
  \end{itemize}

The total logarithmic likelihood of the combined model of RV and astrometry is 
\begin{equation}
    \ln\mathcal{L}=\ln{\mathcal{L}_{\rm rv}}+\ln{\mathcal{L}_{\rm gaia}}.
\end{equation}
Adopting log-uniform priors for time-related parameters, an informative prior for parallax, and uniform priors for other parameters, we launched multiple MCMC samplers to draw posterior samples\cite{haario2006dram}. 
We used two methods to derive the prior of parallax.
Firstly, we performed SED fitting with the PARSEC isochrones (\url{http://stev.oapd.inaf.it/cgi-bin/cmd\_3.1}).
We downloaded the sequences of isochrones with an age step of $\Delta$(log\,t) = 0.005, and collected all the points within the atmospheric parameter range ([$T_{\rm eff} - 100$ K, $T_{\rm eff} + 100$ K], [log$g$ $-$ 0.1, log$g$ $+$ 0.1]) as acceptable models.
For each model, we determined parallax and extinction values by fitting the observed magnitudes ($G$, $BP$, $RP$, $J$, $H$, $K_{\rm S}$, $B$, and $V$).
The distribution of parallax results exhibited two peaks, with Gaussian fittings returning parallax values of 0.68$\pm$0.05 mas and 0.58$\pm$0.05 mas, respectively.
Secondly, we calculated spectroscopic masses using different parallaxes with a step of 0.01 mas, and compared them with the evolutionary mass estimated from the {\it isochrones} code\cite{2015ascl.soft03010M}.
A parallax of 0.62$\pm$0.05 mas/yr was considered a suitable prior, resulting in a spectroscopic mass equal to the evolutionary mass. 
Finally, a moderate prior of 0.6$\pm$0.1 mas/yr was applied by combining the three prior estimates.
Furthermore, we thoroughly explored the option of a priori correction for biases in Gaia DRs, such as parallax zero point and frame rotation. Our investigations revealed that these corrections did not result in any significant changes to the orbital solution.

The best fit of the RV and astrometry models to the data is shown in Extended Data Fig. 2. As shown in the second panel, the orbital phase is well sampled by Gaia observations. With GOST emulating Gaia epochs and refitting to the synthetic data\cite{2023MNRAS.525..607F}, we are able to constrain short period companions without approximating instantaneous position and velocity by catalog position and velocity. The decomposition of the best-fit astrometry model to the five-parameter solutions of Gaia DR2 and DR3 is shown in Extended Data Fig. 3. 
The 1D and 2D posterior distribution of orbital parameters are shown in Supplementary Fig. 3.
The model and data exhibit a close agreement within a 1-$\sigma$ confidence interval for nine astrometric parameters. 
The parameters for the combined RV and astrometry analyses are shown in Extended Data Table 2 and Supplementary Table 2. 
By subtracting the offset ($\Delta \varpi =$ 0.15 mas) from the Gaia DR3 parallax, we derived a new parallax of $0.56^{+0.09}_{-0.09}$ mas for G3425, corresponding to a distance of 1786$^{+342}_{-248}$ pc.
The fitted inclination angle was about $i = 89^{+15}_{-10}$ degrees, suggesting the orbit is close to edge-on. The uncertainty of the inclination angle does not significantly affect $M_2$ because $df(M_2)/di$ is approximately proportional to $M_2*{\rm sin}^2i*{\rm cos}i$.

In addition, as a test, we also applied these steps to Gaia BH2.
We derived the parallax priors of 0.81$\pm$0.1 mas/yr and 0.89$\pm$0.05 mas/yr, and a moderate parallax prior of 0.85$\pm$0.1 mas/yr.
We finally obtained a parallax of 0.84$^{+0.08}_{-0.05}$ mas, which is in agreement with the NTBO solution of $0.86\pm0.02$ mas.
The calculated inclination angle is $40.22_{-3.99}^{+7.43}$ degrees, consistent with previous estimation ($i$ = 34.9$\pm$0.4 degrees)\cite{2023MNRAS.521.4323E}.

\subsection{Stellar properties of the visible star}

G3425 was observed at multiple epochs by LAMOST. 
There are three estimations of stellar parameters available in the LAMOST DR11 low-resolution catalog and five estimations in the LAMOST DR11 medium-resolution catalog. 
Its atmosphere parameters can be estimated following 
\begin{equation} \label{eqweight}
\overline{P} = \frac{\sum_k w_k \cdot P_{k}}{\sum_k w_k}
\end{equation}
and
\begin{equation} \label{eqweight2}
\sigma_w(\overline{P}) = \sqrt{\frac{N}{N-1}\frac{\sum_k w_k \cdot (P_{k} - \overline{P})^2}{\sum_k w_k}},
\end{equation}
where the index $k$ is the $k_{th}$ epoch of the measurements of parameter $P$ (i.e., $T_{\rm eff}$, log$g$, and [Fe/H]) for each star, and the weight $w_k$ is square of SNR of each spectrum according to the $k_{th}$ epoch\cite{2020ApJS..251...15Z}.

Supplementary Table 3 lists the stellar parameters estimated from different methods, including the LAMOST Stellar Parameter Pipeline (LASP)\cite{2015RAA....15.1095L}, the Data-Driven Payne (DD-Payne)\cite{2019ApJS..245...34X}, the Stellar LAbel Machine (SLAM)\cite{2020ApJS..246....9Z}, the Residual Recurrent Neural Network (RRNet)\cite{2022ApJS..261...36X}, and the CYCLE-STARNET method\cite{2023ApJS..266...40W}.
In brief, LASP is mainly based on ULYSS and uses the ELODIE library to determine atmospheric parameters and radial velocities; 
DD-Payne employs a hybrid approach that combines a data-driven method with astrophysical modeling priors, utilizing neural-network spectral interpolation and the fitting algorithm of Payne.
%derives the stellar parameters with a hybrid method that combines the data-driven approach with priors of astrophysical modeling, utilizing neural-network spectral interpolation and the fitting algorithm of Payne; 
SLAM is a data-driven method based on support vector regression and can derive stellar labels across various spectral types; 
RRNet uses a residual module and a recurrent module to synthetically extract spectral information and estimate stellar atmospheric parameters; 
CYCLE-STARNET is an unsupervised domain-adaptation method that combines the data consistency of data-driven methods and the physical interpretability of model-driven methods.
Generally, for K to F stars (4000 $K < T_{\rm eff} < $ 7500 $K$), these methods can give reliable atmospheric estimates\cite{2021RAA....21..292W}. For G3425, most estimations are in good agreement with each other.
Here we used the stellar parameters estimated by LASP using low-resolution observations: $T_{\rm eff} = 4984{\pm 25}$ K, log$g$ $=2.63{\pm 0.05}$, and [Fe/H] $=-0.12{\pm 0.02}$.

We calculated the $V{\rm sin}i$ values from the blue (5300 to 5500 \AA) and red (6500 to 6800 \AA) bands of the LAMOST MRS.
We utilized the PHOENIX template with an effective temperature of 5000 K, a surface gravity of log$g$ $=$ 2.5, and a metallicity of $Z=0$. 
After calculating the $V{\rm sin}i$ for each spectrum, we derived an averaged value using Eqs. \ref{eqweight} and \ref{eqweight2}.
Finally, the $V{\rm sin}i$ calculated from the blue and red bands are $29.36\pm 0.53$ km/s and $28.33\pm 2.21$ km/s, respectively. 
Due to the relatively low resolution of the spectra ($R\approx7500$), the $V{\rm sin}i$ values are only rough estimates.

We obtained the extinction through the star-pair method\cite{2013MNRAS.430.2188Y}, using the three LAMOST LRS observations.
In brief, this method assumes that stars with identical stellar atmosphere parameters exhibit the same absorption line features. 
The intrinsic spectrum of the reddened target star can be deduced from its control pairs or counterparts with the same atmospheric parameters but experiencing either no or well-documented extinction directly derived from the SFD map\cite{1998ApJ...500..525S}. 
The extinction value ($E(B-V) =0.46\pm0.01$) is slightly higher than that obtained from the Pan-STARRS DR1 dust map\cite{2015ApJ...810...25G} ($E(B-V) =0.884 \times {\rm Bayestar19} \approx 0.37$).
Extended Data Fig. 4 shows the position of G3425 in Hertzsprung–Russell diagram. The red and blue stars are calculated with the distances from Gaia DR3 (1442 pc)\cite{2021AJ....161..147B} and our joint fitting (1786 pc), respectively.

The astroARIADNE (\url{https://github.com/jvines/astroARIADNE}), a python module, is designed to fit broadband photometry by using different atmospheric templates, including PHOENIX, BT-Settl, CK04, and Kurucz 1993, etc.
Multi-band magnitudes such as Gaia DR2 ($G$, $G_{\rm BP}$, and $G_{\rm RP}$), 2MASS ($J$, $H$, and $K_{\rm S}$), APASS ($B$, $V$, $g$, $r$, and $i$) and WISE ($W$1 and $W$2) were used. 
The atmospheric parameters, distance, and extinction were also used as input priors.
When the new distance (1786$^{+342}_{-248}$ pc) was utilized, the SED fitting (Extended Data Fig. 5) returns an effective temperature of $4990^{+25}_{-21}$ K, a surface gravity of $2.64^{+0.04}_{-0.05}$, 
 and a metallicity of $-0.13^{+0.02}_{-0.01}$, consistent with spectroscopic estimates, and a radius of $12.77_{-0.41}^{+0.90}\ R_{\odot}$.

We used two methods to estimate the mass of the visible star:

 (1) The {\it isochrones}\cite{2015ascl.soft03010M} module can be used to calculate the evolutionary mass by fitting the photometric or spectroscopic parameters with the MESA models. 
We took the effective temperature $T_{\rm eff}$, surface gravity log$g$, metallicity [Fe/H], multi-band magnitudes ($G$, $G_{\rm BP}$, $G_{\rm RP}$, $J$, $H$, and $K_{\rm S}$), distance and extinction $A_V$ as the input parameters. 
The best-fit model yielded an evolutionary mass of $2.39^{+0.22}_{-0.18}\ M_{\odot}$ and a radius of $12.54_{-1.02}^{+0.85}\ R_{\odot}$.

 (2) We used six magnitudes ($G$, $G_{\rm BP}$, $G_{\rm RP}$, $J$, $H$, $K_{\rm S}$) and the atmospheric parameters to calculate the spectroscopic mass of G3425 following $M = gL_{\rm bol}/(4\pi G\sigma T_{\rm eff}^{4})$.
With the absolute luminosity and magnitude of the sun ($L_{\odot} =$ 3.83$\times$ 10$^{33}$ erg/s; $M_{\odot} =$ 4.74), the bolometric luminosity was calculated with
$L_{\rm bol} = 10^{0.4\times(M_{\odot} - M_{\rm bol})}L_{\odot}$.
%$M_{\odot} - M_{\rm bol} = 2.5\log(L_{\rm bol}/L_{\odot})$. 
%
The bolometric magnitude was calculated following
$m_{\rm \lambda} - M_{\rm bol} = 5{\rm log}d - 5 + A_{\rm \lambda} - BC_{\rm \lambda}$, where $m_{\rm \lambda}$ is the apparent magnitude of each band and $d$ is the distance from our new determination. 
$A_{\rm \lambda}$ is the extinction of each band calculated with $A_{\rm \lambda} = R_{\lambda} \times E(B-V)$, where $E(B-V) = 0.46\pm0.01$ was estimated from the StarPair method\cite{2013MNRAS.430.2188Y} and $R_{\lambda}$ is the extinction coefficient of each band\cite{1999PASP..111...63F}.
$BC_{\lambda}$ is the bolometric correction calculated using the {\it isochrones} package, with $T_{\rm eff}$, log$g$, and [Fe/H] values as the input.
The averaged spectroscopic mass from six bands is about $2.66^{+1.18}_{-0.68} M_{\odot}$, in agreement with the evolutionary mass estimation.

Note that different atmospheric parameters can lead to different mass estimates. Using the parameters from other methods (in Supplementary Table 3), we derived more spectroscopic mass estimates of the giant: $2.36^{+0.99}_{-0.62}$, $1.73^{+0.78}_{-0.49}$, $2.46^{+1.16}_{-0.73}$, $2.86^{+1.22}_{-0.76}$, and $2.31^{+0.97}_{-0.60}\ M_{\odot}$ with LASP/MRS, DD-Payne/LRS, SLAM/MRS, RRNet/MRS, and CYCLE-STAR/MRS, respectively.

Supplementary Table 4 lists the mass estimates of the giant star. 
Considering that the two stars may have interaction during the evolution of the binary system, especially during the early stage of progenitor of the black hole, the evolution of the visible star may deviate slightly from that of a single star. Therefore, the evolutionary mass estimated through isochrone fitting may be inaccurate to some extent. On the contrary, the spectroscopic mass is more reliable as it depends only on the star's current properties and not on its evolutionary history.

\subsection{Optical light curves}

The ASAS-SN $V$- and $g$-band light curves were also used to search for a period, but no clear period was detected.
Supplementary Fig. 4 displays the ASAS-SN $V$- and $g$-band light curves of the system, folded with the orbital period of $\approx$880 days. 
As expected, no obvious ellipsoidal modulation is observed in these light curves. 
The long orbital period suggests this system is a wide binary system, with tiny ellipsoidal deformation of the visible star caused by the tidal force from the unseen companion.

G3425 was observed by TESS in Sectors 43, 44, 45, 71, and 72. However, only in Sector 71 were two faint pixels detected as the counterpart of G3425. The extracted light curve was significantly impacted by instrumental effects and showed no periodic feature.

\subsection{FAST observation}

We performed 40 minutes of radio observations with the Five-hundred-meter Aperture Spherical radio Telescope (FAST), aiming at an exceptionally sensitive radio follow-up for radio pulsation and continuous spectrum search, in two sessions. Session one was on April 21th 2023 from 08:33:00 to 08:53:00 UTC (Coordinated Universal Time), while session two was on June 2th 2023 from 03:52:00 to 04:12:00 UTC. The calibration signal injection time for flux and polarisation calibration was one minute each, at the beginning and end of each observation. For the first session, the observation taken at FAST uses the center beam of the 19-beams L-band receiver, while all beams are used in the second session. The frequency range is from 1.05 to 1.45 GHz with an average system temperature of 25 Kelvin\cite{2020RAA....20...64J}. The observation data was recorded in pulsar search mode and stored in PSRFITS format\cite{2004PASA...21..302H}. During the observational campaign, we conducted two types of data processing:

I. Dedicated and blind search:

Based on the Galactic electron density model NE2001 model\cite{2002astro.ph..7156C} and YMW16 model\cite{2017ApJ...835...29Y}, we firstly estimated the distance $D$=1538 $\pm$ 120 pc corresponding with a dispersion measure (DM) of 46.5--55.5 pc cm$^{-3}$ (NE2001) and 66.6--87.1 pc cm$^{-3}$ (YMW16), and the line of sight maximal Galactic DM (max) = 178.3 pc cm$^{-3}$ (NE2001) and 265.5 pc cm$^{-3}$ (YMW16). Due to model dependence and for the sake of robustness, we created de-dispersed time series for each pseudo-pointing over a range of DMs, from 0--1000 pc cm$^{-3}$ , which is a factor of four larger than the maximum DMs models predicted in the line of sight. For each of the trial DMs, we searched for a periodical signal and first two order acceleration in the power spectrum based on the PRESTO\cite{2001PhDT.......123R} pipeline\cite{2021SCPMA..6429562W}. We checked all the pulsar candidates of SNR $>$ 6 pulse by pulse and removed the narrow-band radio frequency interferences (RFIs).

Both the periodical radio pulsations and single-pulse blind searches were performed for each observing epoch, but resulted in non-detections for all sessions. We calibrated the noise level of the baseline, and then measured the amount of pulsed flux above the baseline, giving the 6$\sigma$ upper limit of flux density measurement of 12 $\pm$ 3 $\mu$Jy in both of the sessions for persistent radio pulsations (assuming a pulse duty cycle of 0.05 - 0.3). The time interval between session (1) and (2) is more than one month, and the effect of interstellar scintillation can be well excluded.

II. Single pulse search:

The above search strategy was continued to de-disperse the time series for single pulse search and flux calibration. We used PRESTO and HEIMDALL\cite{2016MNRAS.460L..30C} softwares. A zero-DM matched filter was applied to mitigate RFI in the blind search. All of the possible candidates were plotted, then be confirmed as RFIs by manual check one by one. No pulsed radio emission with a dispersive signature was detected with an SNR $>$ 6. The upper limit of pulsed radio emission is $\sim$0.015 Jy ms assuming a 1 ms wide burst in terms of integrated flux (fluence).

\subsection{Black holes in the mass gap}

Supplementary Table 5 presents the orbital fitting results of G3425 using different priors, all of which indicate the presence of a black hole with a mass ($\approx$ 3--4 $M_{\odot}$) falling within the mass gap.

The existence of the mass gap has been a long-debated topic since it was initially noticed through the black hole mass distribution discovered via X-ray emission\cite{1998ApJ...499..367B}.
Several alternative theories have been proposed to explain such a gap through revising the process of supernova explosions of massive stars, including rapid convection instability growth\cite{2012ApJ...749...91F,2012ApJ...757...91B}, neutrino-driven explosion\cite{2012ApJ...757...69U}, or failed explosion of red supergiants\cite{2014ApJ...785...28K}.
It was also suggested that low-mass black holes may not exist in binaries because natal kicks above 20--80 km s$^{-1}$ during supernova explosion can disrupt such systems\cite{2021MNRAS.500.1380M}.

On the contrary, recent multi-messenger observations have found a few low-mass balck hole candidates.
Through a correlation between $H_{\alpha}$ trough depth and orbital inclination, the mass of the black hole in GRO J0422+32 was constrained to be $2.7^{+0.5}_{-0.7}$ $M_{\odot}$\cite{2022MNRAS.516.2023C}.
A low-mass black hole candidate ($M = $ 3.3$^{+2.8}_{-0.7}$ $M_{\odot}$) was discovered through RV\cite{2019Sci...366..637T}, although the unseen object could potentially be a close binary consisting of two main-sequence stars\cite{2020Sci...368.3282V}.
In gravitational wave detections by LIGO/Virgo, the compact remnant of GW190425 ($M = 3.4^{+0.3}_{-0.1}\ M_{\odot}$)\cite{2020ApJ...892L...3A} also have a mass within the gap range. 
Recently, a compact object ($M = 2.35^{+0.20}_{-0.18}\ M_{\odot}$) was discovered as a companion to a pulsar through pulsar timing observations\cite{2024Sci...383..275B}. It is thought to be a massive neutron star or a low-mass black hole.
Supernova models, such as slow convection instability growth\cite{2012ApJ...757...91B} or massive fallback to newborn neutron stars\cite{2020ApJ...890...51E}, suggest that black holes can exist in the mass gap.
The discovery of G3425 further provides evidence for the existence of low-mass black holes in noninteracting binaries, which is hard to detect through X-ray emission.
These multi-messenger discoveries hint at a population of dark remnants in the mass gap\cite{2020A&A...636A..20W}, suggesting the gap may not be driven by supernova physics but could be predominantly due to limited statistical data, observational biases, or inaccurate estimates of orbital parameters (i.e., inclination angle)\cite{2012ApJ...757...36K}.

\subsection{Comparison with other known black hole binaries}

To compare the formation scenarios of different black holes in binaries, we 
searched for dynamically confirmed black holes from previous studies, which were discovered by X-ray, RV, and astrometry,
and explored the relations between black hole mass $M_{\rm BH}$, orbital eccentricity $e$, and orbital period $P_{\rm orb}$ (Supplementary Fig. 5).
The X-ray sample includes about twenty Galactic black holes from BlackCAT\cite{2016A&A...587A..61C}, and LMC X-1\cite{2009ApJ...697..573O}, LMC X-3\cite{2014ApJ...794..154O}, M33 X-7\cite{2007Natur.449..872O}, and NGC 300 X-1\cite{2010MNRAS.403L..41C}.
The quiescent black hole sample includes AS 386\cite{2018ApJ...856..158K}, NGC 3201 \#12560\cite{2018MNRAS.475L..15G}, NGC 3201 \#21859\cite{2019A&A...632A...3G},
VFTS 243\cite{2022NatAs...6.1085S}, HD 130298\cite{2022A&A...664A.159M}, Gaia BH1\cite{2023MNRAS.518.1057E}, BH2\cite{2023MNRAS.521.4323E} and BH3\cite{2024A&A...686L...2G}, G3425, and some candidates on debate, such as MWC 656\cite{2014Natur.505..378C} and 2M05215658+4359220\cite{2019Sci...366..637T}.
%and LB-1\cite{2019Natur.575..618L}.}
%
The X-ray systems predominantly concentrate at short periods and small eccentricities.
On the contrary, most black holes identified through RV and astrometry methods have periods ranging from $\approx$10 days to $\approx$1000 days.
Their eccentricities appear to fall into two groups: one with $e <$ 0.1 and the other with $e >$ 0.4.
Supplementary Table 7 lists the parameters of these black holes and candidates.

The short periods of the compact X-ray binaries indicate a necessary common envelope phase. 
However, for low-mass X-ray binaries, which include a massive star ($\gtrsim$ 20 $M_{\odot}$) and a low-mass star ($\lesssim$ 2 $M_{\odot}$), when they enter the common envelope stage, the common envelope is so massive that it's difficult to be expelled and will result in a merger\cite{2003MNRAS.341..385P}.
Addressing this puzzle may involve enhancing the common envelope ejection efficiency\cite{2006MNRAS.369.1152K} or reducing the mass of the black hole progenitors\cite{2016MNRAS.457.1015W} ($\gtrsim$15 $M_{\odot}$).

Wide binaries also face challenges during the common envelope stage. If the secondary has a mass lower than $\approx$3--5 $M_{\odot}$, mass transfer through Roche lobe overflow of the massive primary would trigger the common envelope evolution due to an extreme mass ratio.
The typical effect of the common envelope ejection is reducing orbital separation by a factor of 100--1000, which contradicts the wide orbit of black hole binaries discovered by RV and astrometry.
Only extremely efficient common envelope ejections (e.g., with $\alpha$ larger than 5) can help prevent the binary from shrinking to a close orbit\cite{2022MNRAS.512.5620E}.
Alternatively, these binaries with eccentric wide orbits, like Gaia BH1 and BH2, may form dynamically through an exchange interaction in a dense open cluster\cite{2022MNRAS.512.5620E,2023MNRAS.521.4323E}.
However, such a formation channel is disfavored for G3425 due to its circular orbit.

\subsection{Black hole formation: isolate binary evolution scenario}

The evolutionary scenario of G3425 needs to explain the low mass of the black hole and the wide circular orbit. 

Let's first assume the black hole in this binary system formed from an individual star.
The mass of a black hole depends on three major factors: (i) initial stellar mass; (ii) mass loss/transfer during the star’s life; (iii) the final core-collapse/supernova process. 
The mass loss/transfer and collapse/explosion processes are also responsible for the fate of the binary orbit.

(1) The dynamical capture scenario can be first rejected. 
G3425 is located in the thin disk like Gaia BH1 and BH2. The tidal circulation timescale is much longer than the Hubble time.
To estimate the circularization timescale of a black hole-giant binary, we used the BSE code\cite{2002MNRAS.329..897H} to simulate the orbital evolution. 
The masses of the visible star and the black hole are set to be $2.7 M_{\odot}$ and $3.5 M_{\odot}$, respectively. 
During the evolution, we followed the evolution of the visible star from the zero-age main-sequence phase to the age of $\approx$1\,Gyr when the star evolves off the red-giant stage. Since such a visible star has a convective envelope, we adapt the mechanism involving equilibrium tide with convective damping to deal with the binary orbital evolution.

We varied the initial orbital eccentricities in the range of 0.1--0.5 and the initial orbital periods in the range of 900--1000 days to test their influence on orbital circularization. Our calculations showed that the orbital eccentricity of the binary is nearly unchanged, and the corresponding circularization timescale is always larger than 10$^{11}$\,years over the whole 1\,Gyr. 
Since the visible star (with a radius of $\sim 11\ R_\odot$) is well within its Roche lobe (with a size of $\sim 230\ R_\odot$), tides are not expected to be important independent of the mechanism behind tidal damping.

Besides, the Galactocentric coordinates of Gaia BH1 and BH2 are  ($-$7.70, 0.18, 0.17) kpc and ($-$7.34, $-$0.91, $-$0.08) kpc, respectively; the coordinate of G3425 is ($-$9.89, $-$0.25, 0.14) kpc. Therefore, the location of G3425 is about 2--3 kpc further out, suggesting its environment is likely less dense compared to Gaia BH1 and BH2.

(2) Then, our choice is a small natal kick during the formation of the black hole with negligible mass loss, and simultaneously, the orbit parameters do not change much after this step. 
Given the black hole's low mass, we conducted a series of evolutionary models with the masses of the initial progenitor stars set around 15 $M_{\odot}$, which represents the lower limit for black hole formation\cite{2016ApJ...821...38S}.
The helium cores formed by these models have a mass of about 5--7 $M_{\odot}$, indicating these stars need to lose almost their entire hydrogen envelope before the collapse stage.

We considered several paths for the formation of the system:
  \begin{itemize}
      \item Stable mass transfer through Roche lobe overflow. Not likely. In this scenario, the material is transferred from the black hole's progenitor to the current giant's progenitor, and the orbit generally shrinks during this phase. Unless the stable mass transfer is quite non-conserved and the specific angular momentum is lost from the donor, the orbital period could be increased to around 880 days. The mass ratio of this system seems too large to keep a stable mass transfer, but see Ge et al. for late Hurtzsprung gap stars\cite{2015ApJ...812...40G,2020ApJ...899..132G}. Transferring 10--15 $M_{\odot}$ material is also quite challenging. 
      \item Stable mass loss through stellar wind. Not likely. In this scenario, the star keeps in the main sequence and doesn't become a giant or supergiant. Therefore, the star doesn't fill its Roche lobe, and the orbit keeps expanding. We simulated with MESA for a progenitor with a mass of 18 $M_{\odot}$, and found a 5 $M_{\odot}$ Helium core will be left only if the wind is 10 times stronger than Reimers wind\cite{1975MSRSL...8..369R}. However, our simulation showed the star finally still fills its Roche lobe and the wide orbit can not be maintained. More seriously, it's hard to explain why only this system experiences such an extreme wind.
      Many black holes with small masses should be discovered if the enhanced stellar wind by a factor of 10 is a normal phenomenon.
      On the contrary, most black holes discovered through X-ray contain more massive black holes ($\approx$6--20 $M_{\odot}$).
      \item Unstable mass loss through L2/L3 point. Not likely. Some simulations showed that specific combinations of the mass ratio, orbital eccentricity, and ratio of rotational to orbital angular velocity can lead to lower potential of L2/L3 point than L1 point\cite{2007ApJ...660.1624S}, and the material can be thrown away from the L2/L3 point\cite{2018ApJ...863....5M,2020ApJ...893..106M}, causing the orbit to expand during this phase.
      However, the duration of this stage and the mass loss efficiency remain unknown. Besides, the scenario appears to be a normal phenomenon if it can occur, which hasn't been observed so far.
      \item Common envelope. For binaries with extreme mass ratio, the mass transfer is unstable and it will enter the common envelope phase. We ran a series of simulations using an adiabatic mass-loss model\cite{2010ApJ...717..724G,2015ApJ...812...40G,2020ApJ...899..132G} for the binary evolution. During the common envelope evolution, a most important parameter $\alpha_\mathrm{CE}$ is defined as
      \begin{equation}
	\alpha_\mathrm{CE} \Delta E_\mathrm{orb} = E_\mathrm{bind},
	\label{eq04}
    \end{equation}
      where $\Delta E_\mathrm{orb}$ is the orbital energy change and $E_\mathrm{bind}$ is the binding energy. This formula can also be written as\cite{1984ApJ...277..355W} 
    \begin{equation}
	\alpha_\mathrm{CE} \left( -\frac{\mathrm{G}M_\mathrm{2i}M_\mathrm{1}}{2a_\mathrm{i}} +\frac{\mathrm{G}M_\mathrm{2f}M_\mathrm{1}}{2a_\mathrm{f}} \right) = \frac{\mathrm{G}M_\mathrm{2i}M_\mathrm{2e}}{\lambda R_\mathrm{1i} },
	\label{eq05}
    \end{equation}
    where $\mathrm{G}$ is Newton's gravitational constant, $M$ are the masses, $a$ are the semi-major axes, $R$ is the stellar radius, and $\lambda$ is a dimensionless parameter that reflects the structure of the star. The 1 and 2 subscripts represent the accretor and donor, respectively. The subscripts i and f represent the initial and final state of the common envelope phase. $M_\mathrm{2e}=M_\mathrm{2i}-M_\mathrm{2f}$ is the mass of the common envelope.
    
    The change of the total energy of the donor is calculated by\cite{2010ApJ...717..724G,2022ApJ...933..137G,2024ApJ...961..202G}
    \begin{equation}
    \begin{aligned}
 	\Delta E_2 =-E_\mathrm{bind}= &\int_{0}^{M_{2 \mathrm{f}}}\left(-\frac{G m}{r}+U\right) d m~\\
    &-\int_{0}^{M_{2 \mathrm{i}}}\left(-\frac{G m}{r}+U\right) d m,
     \end{aligned}
     \label{eq06}
    \end{equation}
    where $m$ is the mass, $r$ is the radius and $U$ is its specific internal energy. Combining Eqs. \ref{eq04} to \ref{eq06}, we derived the initial to final separation relation,
    \begin{equation}
	\frac{a_{\mathrm{f}}}{a_{\mathrm{i}}}=\frac{M_{2 \mathrm{f}}}{M_{2 \mathrm{i}}}\left(1+\frac{2 a_{\mathrm{i}} \Delta E_{2}}{\beta_{\mathrm{CE}} G M_{1} M_{2 \mathrm{i}}}\right)^{-1}.
	\label{eq07}
    \end{equation}
    Here, we defined the common envelope efficiency parameter as $\beta_\mathrm{CE}$ to indicate that the binding energy is calculated by Eq. \ref{eq06}, which considers the response and the redistribution of the core and the thin envelope. 
    
    We built a grid of initial stellar models with $M_\mathrm{2i}=$ 13 and 16 $M_{\odot}$, $M_\mathrm{ 1i}$ = 2, 2.5, and 3.2 $M_{\odot}$, and $\beta_\mathrm{CE} =$ 10, 20, 30, and metallicity $Z=0.02$. The numerical simulations showed that it is possible to produce such a system when the common envelope ejection efficiency is set to be higher than 10. That means the binding energy reduces to 1/10 of the original value, or the external energy input is 10 times the orbital energy. Part of the results are listed in Supplementary Table 8.

    The high ejection efficiency suggests a reduced binding energy or external energy input.
    In previous studies, various processes (e.g, nuclear energy\cite{2010MNRAS.406..840P}, jet by an accreting companion\cite{2004NewA....9..399S}) were considered as potential external energy sources.
    However, these processes seem unlikely to provide sufficient energy to remove the massive envelope of G3425.
    We qualitatively suspect a rare supernova explosion occurs during the common envelope stage, leading to the formation of a low-mass black hole and a wide (yet elliptical) orbit.
    Although the typical expelled energy from one type IIn/IIp supernova is about 10$^{51}$ erg, much higher than the binding energy of the common envelope ($\approx$ 10$^{49}$ erg)\cite{2011ApJ...743...49L}, in asymmetrical cases, a portion of the mass may remain gravitationally bound.
    For example, more energy deposition in the equatorial plane might leave some mass bound near the polar directions\cite{2011MNRAS.417.1466K}.
    The thin leftover envelope may help effectively circularize the wide orbit before it totally dissipates.
    \end{itemize}

\subsection{Black hole formation: triple system scenario}

We also considered the triple system scenario, since triple systems are common in the universe, particularly for high-mass early B/O-type primaries, where the fraction of triples can reach several tens of percent\cite{2017ApJS..230...15M}.
We considered three possible formation paths involving the evolution of a triple system, in which the mass-gap black hole is assumed to form from the accretion-/merger-induced collapse of neutron star(s).

(1) The progenitor of G3425 includes the observed giant star as an outermost component and an inner binary containing two massive stars.
For the inner binary, the more massive one evolves first, forming a neutron star.
The system would evolve into a common envelope stage since the other star should have a mass $\gtrsim$ 5 times the neutron-star mass.
The neutron star sinks towards the star's core before the common envelope is ejected and possibly forms a massive Thorne-\.{Z}ytkow object\cite{1995MNRAS.274..485P}.
The present black hole finally formed when a significant fraction of the envelope was accreted onto the core. 

(2) It is also possible that the common envelope was successfully ejected before the merge, and the central unseen object is the merger product of two neutron stars or even still contains two neutron stars.
In the latter case, it would be an exciting candidate of the merger of binary neutron stars (BNSs), which is detectable by gravitational wave observations. 
This scenario cannot be completely excluded so far, and more precise spectroscopic monitoring is needed to search for any short-period RV modulation caused by an inner compact binary.

However, according to binary synthesis\cite{2020A&A...638A..94O}, in the Galactic thin disk, there are about 6600 binaries containing black holes and giant stars and about 2600 black holes formed through mergers of two neutron stars.
Simultaneously, assuming a natal kick velocity of 0 or 40 km/s, the number ratio between BNS systems and systems including a BNS and a tertiary is about 1 to 1 or 3 to 1\cite{2019ApJ...883...23H}, respectively. 
Furthermore, the likelihood of the tertiary being a giant is much lower than that of it being a main-sequence star.
This suggests only dozens to hundreds of triple systems, including a BNS and a giant star, are located in the Galactic disk.
This means the probability of the isolate binary evolution scenario (i.e., a binary containing a black hole and a giant star) is much higher than that of this triple system scenario (i.e., a triple system including inner BNSs and an outer giant star).

(3) G3425 may experience an evolution path similar to the triple system PSR~J0337+1715\cite{2014Natur.505..520R}, consisting of one pulsar and two white dwarfs. PSR~J0337+1715 is suggested to form from a successful ejection of triple-system common envelope\cite{2014ApJ...781L..13T, 2015MNRAS.450.1716S} and has undergone two stable mass transfer stages, resulting in the formation of the millisecond pulsar\cite{2014ApJ...781L..13T}. 
Neutron stars can accrete sufficient material, particularly under super-Eddington accretion conditions, leading to collapse and formation of mass-gap black holes\cite{2022MNRAS.514.1054G}. 
Evolutionary calculations show that intermediate-mass X-ray binaries including neutron stars can evolve into mass-gap black hole+white dwarf binaries with orbital periods $\lesssim 10~{\rm day}$. The unseen object of G3425 may be a close binary comprising a mass-gap black hole and a white dwarf, or their merger product. The nearly circular orbit in G3425 may be the outcome of circularization during the common envelope evolution, which could also be influenced by kicks generated by the supernova and accretion-induced collapse.

\clearpage

 \begin{table*}
 \caption*{\bf Supplementary Table 1 $|$ Barycentric-corrected RV values estimated from LAMOST LRS and MRS and 2.16m telescope. RV$_{\rm b,cor}$ and RV$_{\rm r,cor}$ mean the systemic offsets are corrected. 
 \label{lamost.tab}}
 \centering
\renewcommand{\arraystretch}{0.75}
  \begin{tabular}{cccccc}
 \hline\noalign{\smallskip}
BJD & RV$_{\rm b}$  & RV$_{\rm b,cor}$ & RV$_{\rm r}$  & RV$_{\rm r,cor}$ & Res.\\
(day) & (km/s) & (km/s) & (km/s) & (km/s) & \\
 \hline\noalign{\smallskip}                                        
%2457407.12193 & -13.76$\pm$1.60 & -16.18$\pm$1.67 & -13.76$\pm$1.60 & -16.18$\pm$1.67 & LRS\\
2457407.12193 & -14.76$\pm$2.05 & -16.42$\pm$2.11 & -8.76$\pm$2.45 & -10.47$\pm$2.50 & LRS\\
2458089.19768 & 6.75$\pm$0.45 & 12.46$\pm$0.47 & 7.25$\pm$0.45 & 11.50$\pm$0.47 & MRS\\
2458089.20671 & 7.25$\pm$0.40 & 12.96$\pm$0.42 & 7.75$\pm$0.45 & 12.00$\pm$0.47 & MRS\\
2458089.21643 & 6.75$\pm$0.45 & 12.46$\pm$0.47 & 8.26$\pm$0.45 & 12.51$\pm$0.47 & MRS\\
2458090.18939 & 6.25$\pm$0.35 & 12.01$\pm$0.37 & 7.25$\pm$0.40 & 11.83$\pm$0.42 & MRS\\
2458090.19911 & 6.75$\pm$0.40 & 12.51$\pm$0.42 & 6.75$\pm$0.45 & 11.33$\pm$0.47 & MRS\\
2458090.20814 & 6.25$\pm$0.40 & 12.01$\pm$0.42 & 7.25$\pm$0.40 & 11.83$\pm$0.42 & MRS\\
2458149.01930 & 3.75$\pm$0.35 & 9.91$\pm$0.37 & 5.25$\pm$0.45 & 10.34$\pm$0.47 & MRS\\
2458149.02833 & 3.75$\pm$0.30 & 9.91$\pm$0.32 & 5.75$\pm$0.45 & 10.84$\pm$0.47 & MRS\\
2458149.03805 & 4.25$\pm$0.35 & 10.41$\pm$0.37 & 5.75$\pm$0.45 & 10.84$\pm$0.47 & MRS\\
2458447.25847 & -32.27$\pm$0.35 & -32.46$\pm$0.37 & -31.27$\pm$0.40 & -31.24$\pm$0.42 & MRS\\
2458447.27514 & -31.77$\pm$0.35 & -31.96$\pm$0.37 & -31.27$\pm$0.40 & -31.24$\pm$0.42 & MRS\\
2458447.29111 & -31.77$\pm$0.35 & -31.96$\pm$0.37 & -32.27$\pm$0.35 & -32.24$\pm$0.37 & MRS\\
2458447.30778 & -31.77$\pm$0.30 & -31.96$\pm$0.32 & -31.77$\pm$0.40 & -31.74$\pm$0.42 & MRS\\
2458447.32375 & -31.77$\pm$0.35 & -31.96$\pm$0.37 & -30.77$\pm$0.35 & -30.74$\pm$0.37 & MRS\\
%2458493.10286 & -32.77$\pm$3.80 & -30.20$\pm$3.83 & -32.77$\pm$3.80 & -30.20$\pm$3.83 & LRS\\
2458493.10286 & -35.77$\pm$3.90 & -32.66$\pm$3.93 & -31.77$\pm$2.60 & -33.77$\pm$2.65 & LRS\\
2458558.98226 & -32.77$\pm$0.30 & -32.26$\pm$0.33 & -34.27$\pm$0.40 & -32.51$\pm$0.42 & MRS\\
2458558.99823 & -33.27$\pm$0.30 & -32.76$\pm$0.33 & -34.27$\pm$0.40 & -32.51$\pm$0.42 & MRS\\
2458559.01489 & -33.27$\pm$0.30 & -32.76$\pm$0.33 & -33.77$\pm$0.35 & -32.01$\pm$0.37 & MRS\\
2458828.18409 & 2.25$\pm$0.30 & 2.33$\pm$0.33 & 1.75$\pm$0.35 & 1.63$\pm$0.39 & MRS\\
2458828.20076 & 2.25$\pm$0.30 & 2.33$\pm$0.33 & 1.75$\pm$0.35 & 1.63$\pm$0.39 & MRS\\
2458828.21673 & 2.25$\pm$0.30 & 2.33$\pm$0.33 & 1.75$\pm$0.35 & 1.63$\pm$0.39 & MRS\\
2458828.23271 & 2.25$\pm$0.25 & 2.33$\pm$0.28 & 1.75$\pm$0.35 & 1.63$\pm$0.39 & MRS\\
2460012.97734 & -1.75$\pm$0.65 & -2.29$\pm$0.68 & -1.75$\pm$2.95 & -2.67$\pm$2.96 & MRS\\
2460012.99262 & -0.75$\pm$0.65 & -1.29$\pm$0.68 & -2.25$\pm$2.7 & -3.17$\pm$2.71 & MRS\\
2460013.00790 & -3.25$\pm$0.95 & -3.79$\pm$0.97 & -0.25$\pm$3.0 & -1.17$\pm$3.01 & MRS\\
2460022.00424 & -11.76$\pm$3.25 & -1.86$\pm$3.36 & -5.75$\pm$2.35 & -6.45$\pm$2.45 & LRS\\
%2460022.00424 & -9.76$\pm$2.55 & -1.46$\pm$2.68 & -9.76$\pm$2.55 & -1.46$\pm$2.68 & LRS\\
2460322.03163 & -29.91$\pm$1.81 &--- &-30.31$\pm$3.85 & ---&2.16m \\
2460322.05928 & -33.83$\pm$1.70 &--- &-33.63$\pm$3.71 & ---&2.16m \\
2460322.09096 & -33.91$\pm$1.67 &--- &-32.31$\pm$3.57 & ---&2.16m \\
2460353.08682 & -28.53$\pm$2.94 &--- &-30.94$\pm$4.18 & ---&2.16m \\
2460354.09878 & -26.95$\pm$1.80 &--- &-28.35$\pm$4.15 & ---&2.16m \\
2460371.00370&-27.27$\pm$0.35&-27.9$\pm$0.37&-26.77$\pm$0.45&-27.31$\pm$0.47&MRS \\
2460371.01898&-27.77$\pm$0.35&-28.4$\pm$0.37&-26.77$\pm$0.5&-27.31$\pm$0.52&MRS\\
2460371.03425&-27.27$\pm$0.35&-27.9$\pm$0.37&-26.77$\pm$0.5&-27.31$\pm$0.52&MRS\\
\noalign{\smallskip}\hline
  \end{tabular}
\end{table*}

\begin{table}
\caption*{\bf Supplementary Table 2 $|$ Parameters in the joint fitting to LAMOST RV and Gaia DR2 and DR3 data.}
\label{tab:solution}
\begin{center}
\small
\begin{tabular}{llp{3.5cm}llp{1.3cm}p{1.4cm}}
\hline\hline
Parameter &Unit & Meaning& Value & Prior & $\theta_{\rm min}$ ($\mu$) & $\theta_{\rm max}$ ($\sigma$)\\
\hline
$b^{\rm LAMOST}$&km\,s$^{-1}$&RV offset for LAMOST&$-3.2_{-0.2}^{+0.2}$&Uniform&-4628.0&4628.0\\
$J^{\rm LAMOST}$&km\,s$^{-1}$&RV jitter for LAMOST&$0.1_{-0.1}^{+0.1}$&Uniform&0.0&4628.0\\
$w_{1}^{\rm LAMOST}$&---&Amplitude of component 1 of MA(1) for LAMOST&$0.3_{-0.5}^{+0.4}$&Uniform&-1.0&1.0\\
ln$\tau^{\rm LAMOST}$&---&Logarithmic time scale of MA(1) for LAMOST&$-2.0_{-5.8}^{+6.8}$&Uniform&-11.5&11.5\\
ln$J_{\rm GAIA}$&---&Logarithmic jitter for GAIA&$2.6_{-0.7}^{+0.9}$&Uniform&-11.5&11.5\\
\hline
 \end{tabular}
\end{center}
\end{table}

\begin{table}
\caption*{\bf Supplementary Table 3 $|$ Atmosphere parameters and metallicities of the giant star estimated from different methods. \label{atm_parameters.tab}}
\centering
\setlength{\tabcolsep}{2pt}
 \begin{tabular}{ccccccc}
\hline\noalign{\smallskip}
%Parameter & LASP (LRS) & LASP (MRS) & DD-Payne (LRS) & SLAM (MRS) & RRNet (MRS)\\
Parameter & LASP  & LASP  & DD-Payne  & SLAM  & RRNet & CYCLE-STARNET \\
 & (LRS) &  (MRS) &  (LRS) &  (MRS) &  (MRS) & (MRS)\\
\hline\noalign{\smallskip}
$T_{\rm eff}$ (K) & 4984$\pm$25 & 4859$\pm$14 & 4875$\pm$23 & 4990$\pm$43 & 4715$\pm$15 & 4881$\pm$28 \\
${\rm log}g$ & 2.63$\pm$0.05 & 2.54$\pm$0.03 & 2.41$\pm$0.06 & 2.60$\pm$0.08 & 2.57$\pm$0.03 & 2.54$\pm$0.01 \\
${\rm [Fe/H]}$ & -0.12$\pm$0.02 & -0.33$\pm$0.05 & -0.18$\pm$0.02 & --- & -0.15$\pm$0.02 & -0.20$\pm$0.02 \\
${\rm [\alpha/Fe]}$ & --- & --- & 0.00$\pm$0.01 & --- & --- & --- \\
${\rm [Na/Fe]}$ & --- & --- & 0.10$\pm$0.06 & --- & --- & 0.02$\pm$0.07 \\
${\rm [Mg/Fe]}$ & --- & 0.02$\pm$0.02 & -0.04$\pm$0.03 & --- & 0.07$\pm$0.01 & 0.08$\pm$0.05 \\
${\rm [Al/Fe]}$ & --- & 0.02$\pm$0.01 & 0.01$\pm$0.07 & --- & --- & --- \\
${\rm [Si/Fe]}$ & --- & 0.02$\pm$0.01 & 0.02$\pm$0.04 & --- & 0.11$\pm$0.01 & 0.05$\pm$0.01 \\
${\rm [Ca/Fe]}$ & --- & 0.02$\pm$0.01 & -0.02$\pm$0.02 & --- & -0.03$\pm$0.01 & 0.02$\pm$0.01 \\
${\rm [O/Fe]}$ & --- & -0.01$\pm$0.01 & -0.05$\pm$0.06 & --- & --- & --- \\
${\rm [Ti/Fe]}$ & --- & -0.01$\pm$0.01 & 0.07$\pm$0.03 & --- & 0.01$\pm$0.01 & 0.04$\pm$0.02 \\
${\rm [Cr/Fe]}$ & --- & -0.04$\pm$0.03 & 0.07$\pm$0.04 & --- & -0.01$\pm$0.01 & 0.04$\pm$0.03 \\
${\rm [Mn/Fe]}$ & --- & --- & 0.07$\pm$0.04 & --- & -0.16$\pm$0.01 & -0.07$\pm$0.09 \\
${\rm [Co/Fe]}$ & --- & --- & -0.06$\pm$0.03 & --- & --- & 0.12$\pm$0.11 \\
${\rm [Ni/Fe]}$ & --- & 0.01$\pm$0.01 & -0.03$\pm$0.02 & --- & 0.03$\pm$0.01 & 0.06$\pm$0.02 \\
${\rm [Cu/Fe]}$ & --- & -0.07$\pm$0.05 & 0.07$\pm$0.18 & --- & -0.20$\pm$0.02 & 0.09$\pm$0.02 \\
${\rm [Ba/Fe]}$ & --- & --- & 0.62$\pm$0.10 & --- & --- & --- \\
\noalign{\smallskip}\hline
\end{tabular}
\end{table}

\begin{table}
\caption*{\bf Supplementary Table 4 $|$ Spectroscopic mass and evolutionary mass estimations of the giant star using different distances.\label{gra_mass.tab}}
\centering
\setlength{\tabcolsep}{10pt}
 \begin{tabular}{ccccc}
\hline\noalign{\smallskip}
Distance & $M_{\rm 1,spec}$ & $R_{\rm 1,spec}$ & $M_{\rm 1,evo}$ &  $R_{\rm 1,evo}$\\
(pc) & ($M_{\odot}$) & ($R_{\odot}$) & ($M_{\odot}$) & ($R_{\odot}$) \\
\hline\noalign{\smallskip}
$1442\pm 100$ (Gaia DR3) & $1.72\pm 0.29$ & $10.52\pm 0.68$ & $1.72^{+0.15}_{-0.21}$ & $11.05^{+0.20}_{-0.29}$\\
$1786_{-248}^{+342}$ (our fit) & $2.66^{+1.18}_{-0.68}$ & $12.97_{-1.77}^{+2.43}$ & $2.39^{+0.22}_{-0.18}$ & $12.54_{-1.02}^{+0.85}$\\
\noalign{\smallskip}\hline
\end{tabular}
\end{table}

 \begin{table*}
 \caption*{\bf Supplementary Table 5 $|$ RV and astrometric fittings of Gaia BH2 and G3425 using different parallax priors. \label{comfit.tab}}
 \begin{center}
\renewcommand{\arraystretch}{0.9}
  \begin{tabular}{cccccc}
 \hline\noalign{\smallskip}
Plx prior & Plx  & $i$ & f(m)  & M1 & M2 \\
(mas) & (mas) & (deg) & ($M_{\odot}$) & ($M_{\odot}$) & ($M_{\odot}$) \\
 \hline\noalign{\smallskip}       
\multicolumn{6}{c}{Gaia BH2}
\\\hline
\multirow{2}*{$0.81\pm0.10^a$} & \multirow{2}*{$0.82_{-0.04}^{+0.06}$} & \multirow{2}*{$39.33_{-3.37}^{+5.78}$} & \multirow{2}*{$1.35_{-0.02}^{+0.01}$} & $1.22_{-0.54}^{+0.93}$ & $7.33_{-1.88}^{+1.91}$ \\
 &  &  &  & $1.21_{-0.03}^{+0.05}$ & $7.23_{-1.58}^{+1.38}$ \\
\multirow{2}*{$0.89\pm0.05^b$} & \multirow{2}*{$0.87_{-0.05}^{+0.05}$} & \multirow{2}*{$43.11_{-5.37}^{+5.44}$} & \multirow{2}*{$1.36_{-0.02}^{+0.01}$} & $1.09_{-0.46}^{+0.82}$ & $6.13_{-1.48}^{+1.93}$ \\
 &  &  &  & $0.90_{-0.03}^{+0.05}$ & $5.73_{-1.09}^{+1.64}$ \\
\multirow{2}*{$0.85\pm0.10^c$} & \multirow{2}*{$0.84_{-0.05}^{+0.08}$} & \multirow{2}*{$40.22_{-3.99}^{+7.43}$} & \multirow{2}*{$1.35_{-0.02}^{+0.02}$} & $1.12_{-0.49}^{+0.89}$ & $6.94_{-2.05}^{+2.02}$ \\
 &  &  &  & $0.99_{-0.06}^{+0.09}$ & $6.63_{-1.73}^{+1.53}$ \\
\hline
\multicolumn{6}{c}{G3425}
\\\hline
\multirow{2}*{$0.58\pm0.05^a$} & \multirow{2}*{$0.53_{-0.04}^{+0.04}$} & \multirow{2}*{$88.34_{-4.36}^{+5.24}$} & \multirow{2}*{$1.15_{-0.01}^{+0.01}$} & $3.04_{-0.48}^{+0.60}$ & $3.79_{-0.30}^{+0.35}$ \\
 &  &  &  & $2.52_{-0.12}^{+0.21}$ & $3.48_{-0.09}^{+0.13}$ \\
  \multirow{2}*{$0.68\pm0.05^a$} & \multirow{2}*{$0.66_{-0.05}^{+0.04}$} & \multirow{2}*{$85.68_{-9.12}^{+11.77}$} & \multirow{2}*{$1.17_{-0.02}^{+0.03}$} & $1.90_{-0.29}^{+0.41}$ & $3.19_{-0.25}^{+0.31}$ \\
 &  &  &  & $2.19_{-0.18}^{+0.15}$ & $3.36_{-0.15}^{+0.19}$ \\
\multirow{2}*{$0.62\pm0.05^b$} & \multirow{2}*{$0.58_{-0.04}^{+0.04}$} & \multirow{2}*{$84.92_{-6.14}^{+5.78}$} & \multirow{2}*{$1.16_{-0.02}^{+0.01}$} & $2.49_{-0.42}^{+0.54}$ & $3.50_{-0.29}^{+0.34}$ \\
 &  &  &  & $2.39_{-0.14}^{+0.16}$ & $3.43_{-0.11}^{+0.14}$ \\
 \multirow{2}*{$0.60\pm0.10^c$} & \multirow{2}*{$0.56_{-0.09}^{+0.09}$} & \multirow{2}*{$89.30_{-10.08}^{+15.48}$} & \multirow{2}*{$1.10_{-0.03}^{+0.02}$} & $2.66_{-0.68}^{+1.18}$ & $3.58_{-0.47}^{+0.80}$ \\
 &  &  &  & $2.40_{-0.16}^{+0.21}$ & $3.42_{-0.11}^{+0.16}$ \\
\noalign{\smallskip}\hline
\end{tabular}
\end{center}
NOTE. $^a$ means the parallax prior was derived from SED fitting; $^b$ means the parallax prior was derived by equaling the spectroscopic mass with the gravitational mass; $^c$ means a moderate prior, which is finally used. Masses in two rows represent spectroscopic and evolutionary masses, respectively.
\end{table*}

\begin{table*}
 \caption*{\bf Supplementary Table 6 $|$ Results of spectral disentangling test with different mass ratios. "\ding{52}\ding{52}" means the spectra are well disentangled, "\ding{52}" means the spectra can be disentangled but in low significance, and "\ding{53}" means the spectra can not be disentangled. \label{test_dis.tab}}
 \begin{center}
 \setlength{\tabcolsep}{2mm}
\renewcommand{\arraystretch}{0.9}
  \begin{tabular}{cccccccc}
 \hline\noalign{\smallskip}
\multicolumn{8}{c}{Test set}\\
\hline
 & $q$ & $0.70$  & $0.75$ & $0.80$ & $1.00$ & $1.20$ & $1.40$ \\
 & $M_1$ ($M_{\odot}$)&  2.7    &  2.7    &  2.7    &   2.7   &  2.7    &  2.7   \\
 & $M_2$ ($M_{\odot}$)&  3.9    &  3.6    &  3.4    &   2.7   &  2.3    &  1.9   \\
 & $L_2/L_1$ & 0.33 & 0.44 & 0.61 & 1.44& 2.28& 7.04 \\
 \hline\noalign{\smallskip}       
\multicolumn{8}{c}{Atmosphere parameters of BT-COND template}
\\\hline
& $T_{\rm eff}$ ($K$) & 14000 & 13000 & 12500 & 10400 & 9800 & 8200 \\
& ${\rm log}g$ & 4.00 & 4.00 & 4.00 & 4.00 & 4.00 & 4.00 \\
& ${\rm [Fe/H]}$ & 0 & 0 & 0 & 0 & 0 & 0 \\
\hline
\multicolumn{8}{c}{Test results}
\\\hline
 \multirow{4}*{Blue band} & 10 (km/s) & \ding{52}\ding{52} & \ding{52}\ding{52} & \ding{52}\ding{52} & \ding{52}\ding{52} & \ding{52}\ding{52} & \ding{52}\ding{52} \\
& 50 (km/s) & \ding{52}\ding{52} & \ding{52}\ding{52} & \ding{52}\ding{52} & \ding{52}\ding{52} & \ding{52}\ding{52} & \ding{52}\ding{52} \\
& 100 (km/s) & \ding{52}\ding{52} & \ding{52}\ding{52} & \ding{52} & \ding{52} & \ding{52} & \ding{52} \\
& 150 (km/s) & \ding{52} & \ding{52} & \ding{53} & \ding{53} & \ding{53} & \ding{53} \\
\hline
 \multirow{4}*{Red band} & 10 (km/s) & \ding{52}\ding{52} & \ding{52}\ding{52} & \ding{52}\ding{52} & \ding{52}\ding{52} & \ding{52}\ding{52} & \ding{52}\ding{52} \\
& 50 (km/s) & \ding{52}\ding{52} & \ding{52}\ding{52} & \ding{52}\ding{52} & \ding{52}\ding{52} & \ding{52}\ding{52} & \ding{52}\ding{52} \\
& 100 (km/s) & \ding{52}\ding{52} & \ding{52}\ding{52} & \ding{52}\ding{52} & \ding{52}\ding{52} & \ding{52}\ding{52} & \ding{52}\ding{52} \\
& 150 (km/s) & \ding{52}\ding{52} & \ding{52}\ding{52} & \ding{52}\ding{52} & \ding{52}\ding{52} & \ding{52}\ding{52} & \ding{52}\ding{52} \\
\noalign{\smallskip}\hline
\end{tabular}
\end{center}
\end{table*}

\begin{table*}
\caption*{\bf Supplementary Table 7 $|$ Database of black holes and candidates for comparisons. \label{bhs.tab}}
\centering
\small
\renewcommand{\arraystretch}{0.85}
\setlength{\tabcolsep}{2mm}
\begin{center}
\begin{tabular}{lccccc}
\hline\noalign{\smallskip}
Name & $M_{\rm 2}$ ($M_{\odot}$) & $P$ (day) & $e$ & method & References \\
\hline\noalign{\smallskip}
GRO J0422+32 & $2.70^{+0.70}_{-0.50}$ & $0.21^{+0.01}_{-0.01}$ & 0 & X-ray & \cite{2022MNRAS.516.2023C} \\
3A 0620-003 & $6.60^{+0.25}_{-0.25}$ & $0.32^{+0.01}_{-0.01}$ & 0 & X-ray & \cite{2010ApJ...710.1127C} \\
GRS 1009-45 & $\geq4.40$ & $0.29^{+0.01}_{-0.01}$ & 0 & X-ray & \cite{1999PASP..111..969F} \\
XTE J1118+480 & $7.55^{+0.65}_{-0.65}$ & $0.17^{+0.01}_{-0.01}$ & 0 & X-ray & \cite{2013AJ....145...21K} \\
GS 1124-684 & $11.00^{+2.10}_{-1.40}$ & $0.43^{+0.01}_{-0.01}$ & 0 & X-ray & \cite{2016ApJ...825...46W} \\
GS 1354-64 & $\geq7.60$ & $2.54^{+0.01}_{-0.01}$ & 0 & X-ray & \cite{2009ApJS..181..238C} \\
4U 1543-475 & $9.40^{+1.00}_{-1.00}$ & $1.12^{+0.01}_{-0.01}$ & 0 & X-ray & \cite{2003IAUS..212..365O} \\
XTE J1550-564 & $11.70^{+3.90}_{-3.90}$ & $1.54^{+0.01}_{-0.01}$ & 0 & X-ray & \cite{2011ApJ...730...75O} \\
XTE J1650-500 & $\leq7.30$ & $0.32^{+0.01}_{-0.01}$ & 0 & X-ray & \cite{2004ApJ...616..376O} \\
GRO J1655-40 & $6.00^{+0.40}_{-0.40}$ & $2.62^{+0.01}_{-0.01}$ & 0 & X-ray & \cite{2003MNRAS.339.1031S} \\
GX 339-4 & $10.09^{+1.81}_{-1.81}$ & $1.76^{+0.01}_{-0.01}$ & 0 & X-ray & \cite{2017ApJ...846..132H} \\
H 1705-250 & $6.40^{+1.50}_{-1.50}$ & $0.52^{+0.01}_{-0.01}$ & 0 & X-ray & \cite{1997AJ....114.1170H} \\
SAX J1819.3-2525 & $6.40^{+0.60}_{-0.60}$ & $2.82^{+0.01}_{-0.01}$ & 0 & X-ray & \cite{2014ApJ...784....2M} \\
MAXI J1820+070 & $6.90^{+1.20}_{-1.20}$ & $0.69^{+0.01}_{-0.01}$ & 0 & X-ray & \cite{2020ApJ...893L..37T} \\
XTE J1859+226 & $7.80^{+1.90}_{-1.90}$ & $0.28^{+0.01}_{-0.01}$ & 0 & X-ray & \cite{2022MNRAS.517.1476Y} \\
GRS 1915+105 & $11.20^{+2.00}_{-2.00}$ & $33.83^{+0.17}_{-0.17}$ & 0 & X-ray & \cite{2014ApJ...796....2R} \\
GS 2000+251 & $7.15^{+1.65}_{-1.65}$ & $0.34^{+0.01}_{-0.01}$ & 0 & X-ray & \cite{2004AJ....127..481I} \\
GS 2023+338 & $9.00^{+0.20}_{-0.60}$ & $6.47^{+0.01}_{-0.01}$ & 0 & X-ray & \cite{2010ApJ...716.1105K} \\
Cyg X-1 & $21.20^{+2.20}_{-2.20}$ & $5.60^{+0.01}_{-0.01}$ & $0.019^{+0.002}_{-0.002}$ & X-ray & \cite{2021Sci...371.1046M} \\
MAXI J1305-704 & $8.90^{+1.60}_{-1.00}$ & $0.40^{+0.01}_{-0.01}$ & 0 & X-ray & \cite{2021MNRAS.506..581M} \\
M33 X7 & $15.65^{+1.45}_{-1.45}$ & $3.45^{+0.01}_{-0.01}$ & $0.019^{+0.008}_{-0.008}$ & X-ray & \cite{2007Natur.449..872O} \\
LMC X-3 & $6.98^{+0.56}_{-0.56}$ & $1.71^{+0.01}_{-0.01}$ & 0 & X-ray & \cite{2014ApJ...794..154O} \\
LMC X-1 & $10.91^{+1.41}_{-1.41}$ & $3.91^{+0.01}_{-0.01}$ & 0 & X-ray & \cite{2009ApJ...697..573O} \\
NGC 300 X-1 & $14.50^{+3.00}_{-2.50}$ & $1.35^{+0.01}_{-0.01}$ & 0 & X-ray & \cite{2010MNRAS.403L..41C} \\
AS 386 & $\geq7.00$ & $131.27^{+0.09}_{-0.09}$ & 0 & RV & \cite{2018ApJ...856..158K} \\
GC NGC3201 \#12560 & $\geq4.36$ & $166.88^{+0.71}_{-0.63}$ & $0.610^{+0.020}_{-0.020}$ & RV & \cite{2018MNRAS.475L..15G} \\
GC NGC3201 \#21859 & $\geq7.68$ & $2.24^{+0.01}_{-0.01}$ & $0.070^{+0.040}_{-0.040}$ & RV & \cite{2019A&A...632A...3G} \\
VFTS 243 & $\geq8.70$ & $10.40^{+0.01}_{-0.01}$ & $0.017^{+0.010}_{-0.010}$ & RV & \cite{2022NatAs...6.1085S} \\
HD 130298 & $8.80^{+2.50}_{-2.50}$ & $14.63^{+0.01}_{-0.01}$ & $0.457^{+0.007}_{-0.007}$ & RV & \cite{2022A&A...664A.159M} \\
Gaia BH1 & $9.78^{+0.18}_{-0.18}$ & $185.59^{+0.05}_{-0.05}$ & $0.454^{+0.005}_{-0.005}$ & RV+Astrometry & \cite{2023MNRAS.518.1057E} \\
Gaia BH2 & $8.93^{+0.33}_{-0.33}$ & $1276.70^{+0.60}_{-0.60}$ & $0.518^{+0.002}_{-0.002}$ & RV+Astrometry & \cite{2023MNRAS.521.4323E} \\
Gaia BH3 & $32.7\pm0.82$ & $4253.1\pm98.5$ & $0.7291\pm0.0048$ & RV+Astrometry & \cite{2024A&A...686L...2G} \\
MWC 656 & $5.35^{+1.55}_{-1.55}$ & $60.37^{+0.04}_{-0.04}$ & $0.080^{+0.060}_{-0.060}$ & RV & \cite{2014Natur.505..378C} \\
2M05215658+4359220 & $3.30^{+2.80}_{-0.70}$ & $82.20^{+2.50}_{-2.50}$ & $0.005^{+0.003}_{-0.003}$ & RV & \cite{2019Sci...366..637T} \\
PSR  J0514-4002E & $2.35^{+0.20}_{-0.18}$ & 7.4479 & 0.7079 & Pulsar timing & \cite{2024Sci...383..275B} \\
\noalign{\smallskip}\hline
\end{tabular}
\end{center}
\end{table*}

\begin{table*}
 \caption*{\bf Supplementary Table 8 $|$ Simulations using an adiabatic mass-loss model for binary evolution during the common envelope stage. The models with final orbital periods longer than 500 days are shown. 
 \label{form.tab}}
 \begin{center}
\renewcommand{\arraystretch}{0.85}
  \begin{tabular}{ccccccccccc}
 \hline\noalign{\smallskip}
$\beta$ & $M_{\rm 2i}$  & $M_{\rm 2f}$ & $R_{\rm 2i}$  & $R_{\rm 2f}$ & $M_{\rm 1}$  & $R_{\rm 1}$  &  $P_{\rm i}$  & $P_{\rm f}$ & $a_{\rm i}$  & $a_{\rm f}$ \\
($\alpha_{\rm CE}$) & ($M_{\odot}$)  & ($M_{\odot}$) & ($R_{\odot}$)  & ($R_{\odot}$) & ($M_{\odot}$)  & ($R_{\odot}$)  &  (days)  & (days) & ($a_{\odot}$)  & ($a_{\odot}$) \\
 \hline\noalign{\smallskip}                  
30 & 16 &6.92 & 889 & 304 & 3.20 & 2.12 & 1866 & 647 & 1707 & 681 \\
30 & 16 &6.92 & 889 & 301 & 2.50 & 1.82 & 1791 & 614 & 1641 & 642\\
30 & 16 &6.91 & 889 & 297 & 2.00 & 1.61 & 1727 & 581 & 1587 & 607\\
30 & 16 &6.62 & 1048 & 354 & 3.20 & 2.12 & 2387 & 837 & 2012 & 800\\
30 & 16 &6.62 & 1048 & 354 & 2.50 & 1.82 & 2291 & 808 & 1934 & 762\\
30 & 16 &6.62 & 1048 & 354 & 2.00 & 1.61 & 2209 & 779 & 1870 & 730\\
30 & 16 &6.52 & 1062 & 354 & 3.20 & 2.12 & 2435 & 845 & 2039 & 802\\
30 & 16 &6.52 & 1062 & 355 & 2.50 & 1.82 & 2338 & 817 & 1960 & 765\\
30 & 16 &6.52 & 1062 & 355 & 2.00 & 1.61 & 2254 & 789 & 1895 & 734\\
20 & 16 &6.90 & 889 & 291 & 3.20 & 2.12 & 1866 & 606 & 1707 & 651\\
20 & 16 &6.89 & 889 & 285 & 2.50 & 1.82 & 1791 & 568 & 1641 & 609\\
20 & 16 &6.88 & 889 & 279 & 2.00 & 1.61 & 1727 & 531 & 1587 & 571\\
20 & 16 &6.61 & 1048 & 347 & 3.20 & 2.12 & 2387 & 812 & 2012 & 784\\
20 & 16 &6.61 & 1048 & 346 & 2.50 & 1.82 & 2291 & 779 & 1934 & 744\\
20 & 16 &6.61 & 1048 & 343 & 2.00 & 1.61 & 2209 & 745 & 1870 & 709\\
20 & 16 &6.51 & 1062 & 349 & 3.20 & 2.12 & 2435 & 825 & 2039 & 789\\
20 & 16 &6.51 & 1062 & 348 & 2.50 & 1.82 & 2338 & 793 & 1960 & 750\\
20 & 16 &6.51 & 1062 & 346 & 2.00 & 1.61 & 2254 & 762 & 1895 & 717\\
10 & 16 &6.84 & 889 & 257 & 3.20 & 2.12 & 1866 & 505 & 1707 & 575\\
10 & 16 &6.59 & 1048 & 327 & 3.20 & 2.12 & 2387 & 742 & 2012 & 738\\
10 & 16 &6.58 & 1048 & 321 & 2.50 & 1.82 & 2291 & 698 & 1934 & 691\\
10 & 16 &6.58 & 1048 & 314 & 2.00 & 1.61 & 2209 & 654 & 1870 & 649\\
10 & 16 &6.50 & 1062 & 332 & 3.20 & 2.12 & 2435 & 769 & 2039 & 753\\
10 & 16 &6.49 & 1062 & 327 & 2.50 & 1.82 & 2338 & 723 & 1960 & 705\\
10 & 16 &6.48 & 1062 & 318 & 2.00 & 1.61 & 2254 & 674 & 1895 & 659\\
\noalign{\smallskip}\hline
  \end{tabular}
  \end{center}
NOTE. The subscripts ``1" and ``2" represent the accretor (i.e., the progenitor of current giant star) and donor (i.e., the progenitor of the black hole), respectively. The subscripts ``i" and ``f" represent the initial and final state of the common envelope phase, respectively.
\end{table*}

\clearpage

\begin{figure*}
    \center
    \includegraphics[width=1\textwidth]{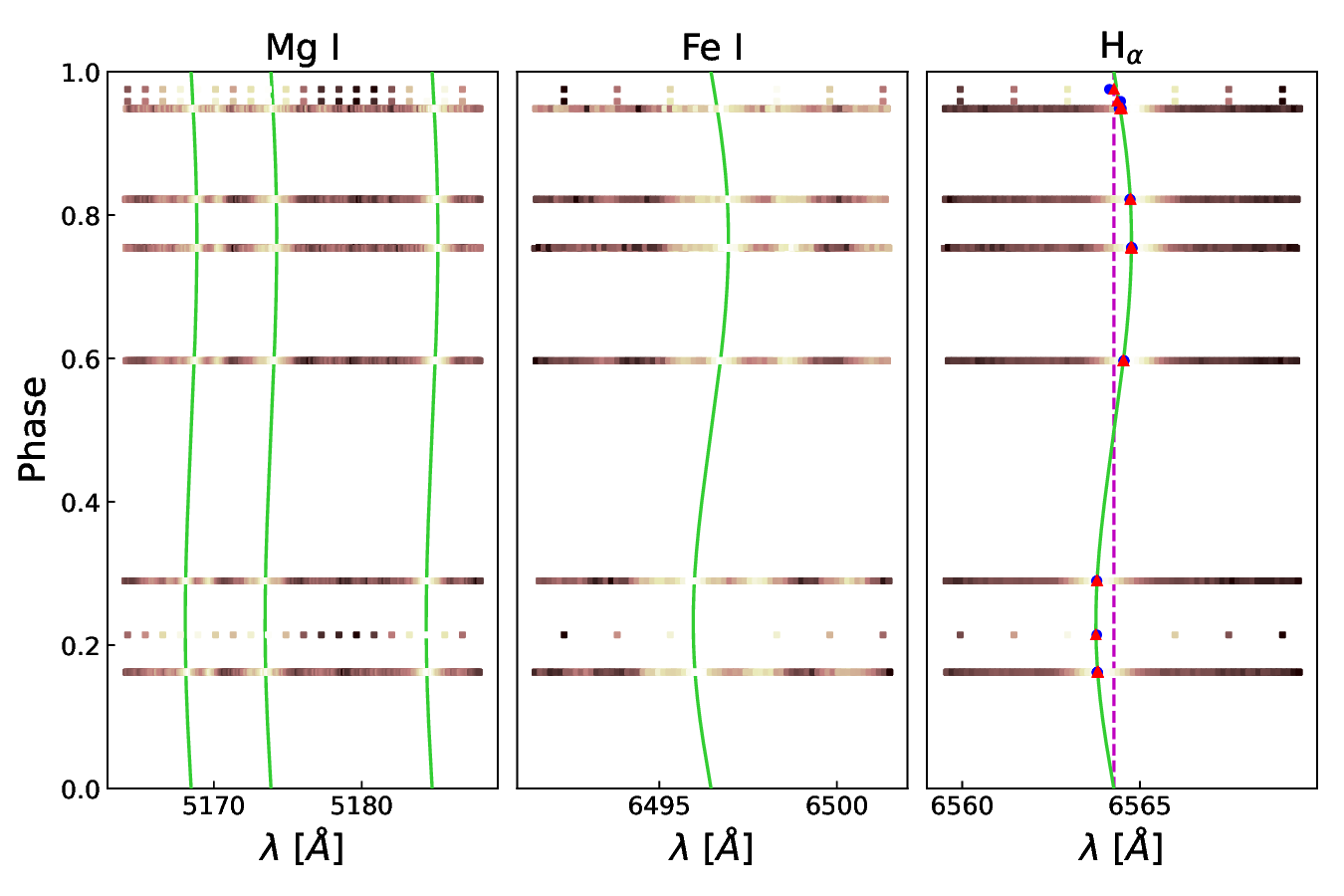}
    \caption*{{\bf Supplementary Fig. 1 $|$ Phase-folded Mg I, Fe I and H$_{\alpha}$ lines.} The bright parts in each spectrum mark these absorption lines. The continuous spectra represent LAMOST medium-resolution observations, while the spectra with separated points represent LAMOST low-resolution observations. The blue dots and red triangles mark the RVs measured from the blue and red bands of the LAMOST spectra, respectively.}
    \label{spec2d.fig}
\end{figure*}

\begin{figure*}
    \center
    \includegraphics[width=0.6\textwidth]{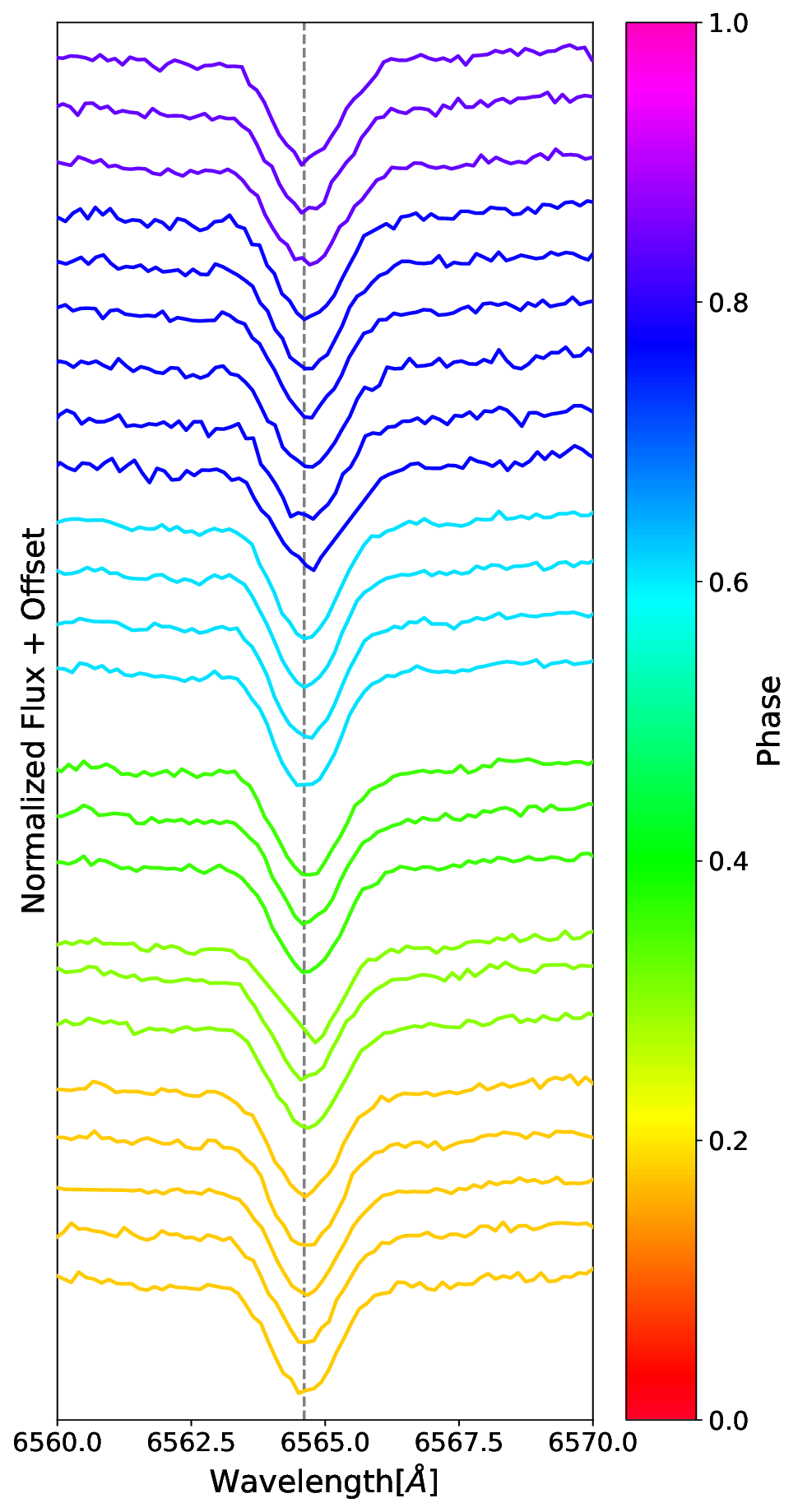}
    \caption*{{\bf Supplementary Fig. 2 $|$ $H_{\alpha}$ line profiles from LAMOST MRS observations plotted in different phases.} The observations with BJD$\approx$2460013 are not plotted because the $H_{\alpha}$ lines were over-subtracted due to an overestimated sky background, which was contaminated by nebula emissions that day.}
    \label{Ha.fig}
\end{figure*}

\begin{figure}
  \centering
  \includegraphics[scale=0.4]{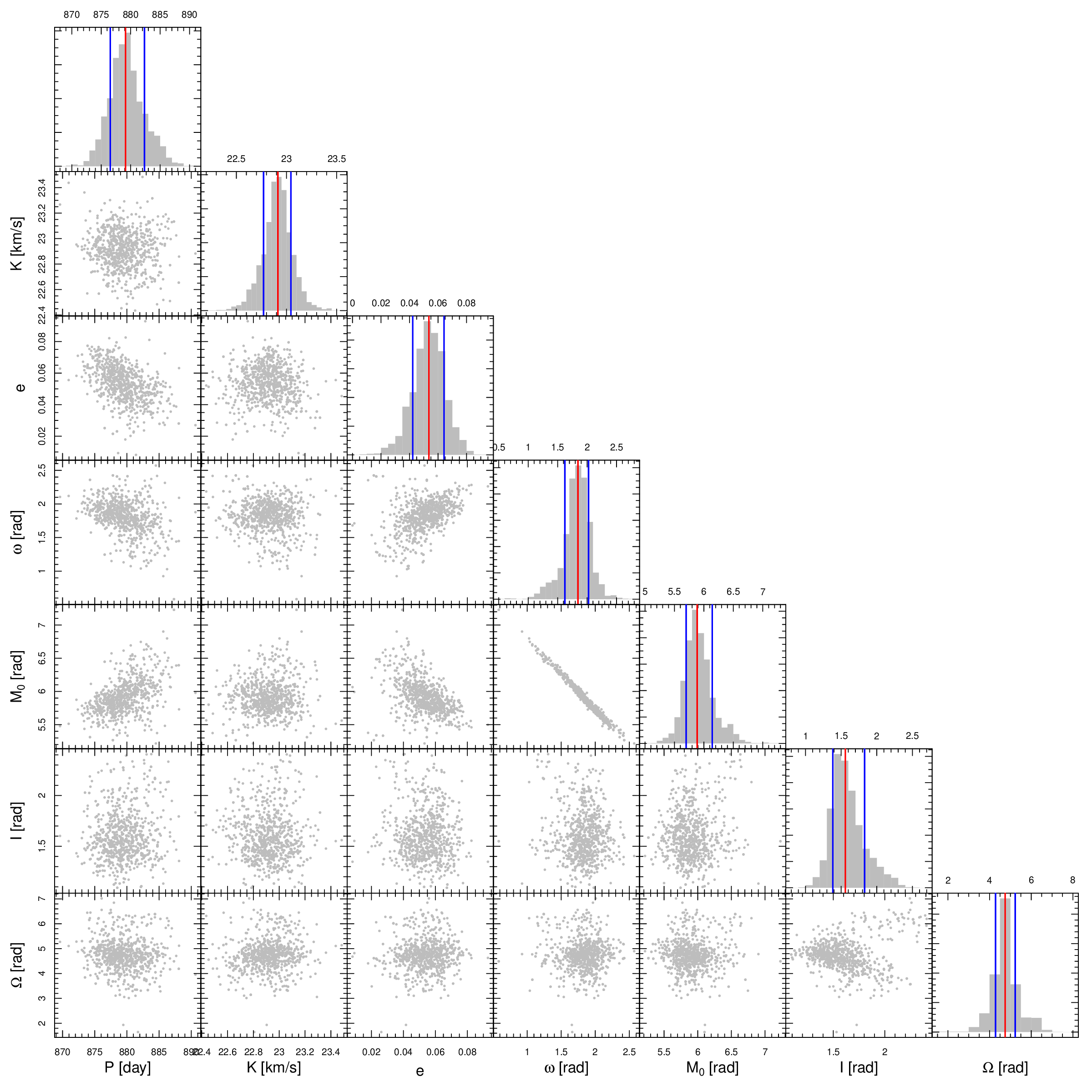}
  \caption*{{\bf Supplementary Fig. 3 $|$ 1D and 2D posterior distribution of orbital parameters for G3425 from the joint fitting.} Each histogram features blue lines denoting the 1$\sigma$ confidence intervals, with the red line indicating the median of the posterior distribution. Posterior samples for each parameter pair are represented by 1000 grey dots.}
  \label{fig:corner}
\end{figure}

\begin{figure*}
    \center
    \includegraphics[width=1\textwidth]{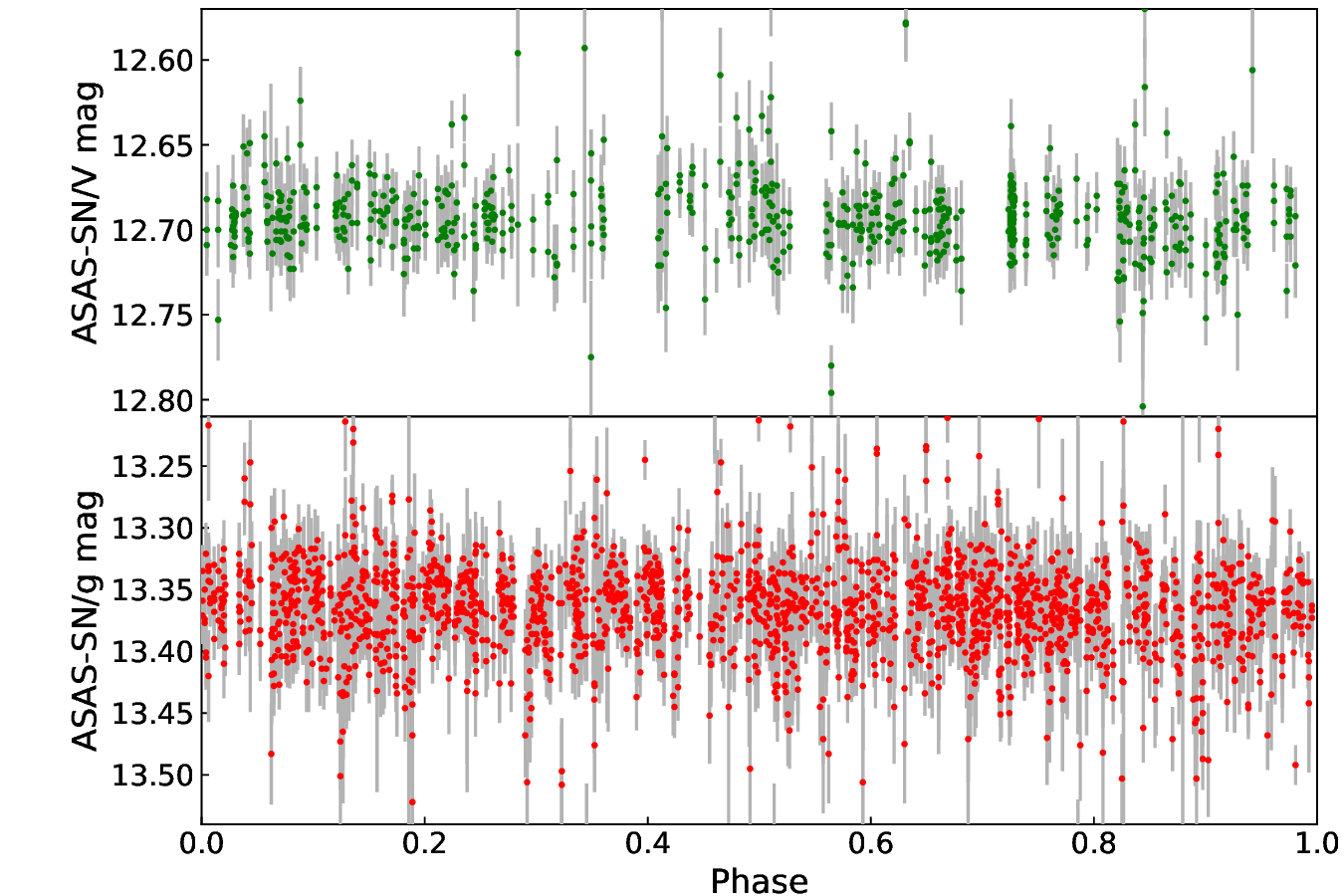}
    \caption*{{\bf Supplementary Fig. 4 $|$ Folded ASAS-SN $V$-band and $g$-band light curves of G3425 with a period of $\approx$880 days.} The error bars represent 1$\sigma$ uncertainties.}
    \label{lcs.fig}
\end{figure*}

\begin{figure*}
    \center
    \includegraphics[width=1\textwidth]{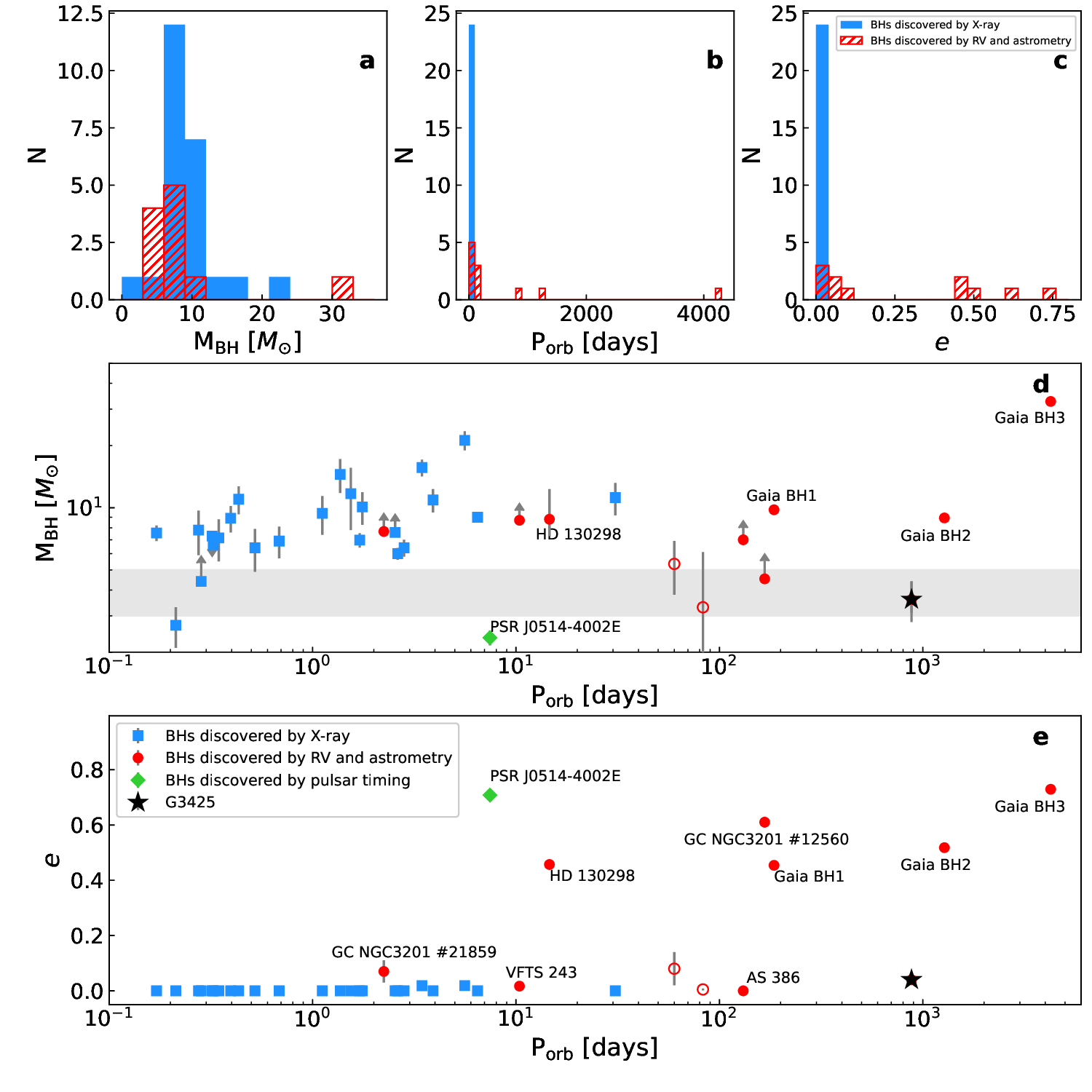}
    \caption*{{\bf Supplementary Fig. 5 $|$ Comparison of G3425 to other known black holes and candidates.}
    Panel a: mass distribution of black holes discovered by X-ray, RV and astrometry. 
    Panel b: orbital period distribution. 
    Panel c: eccentricity distribution. 
    Panel d: black hole mass versus orbital period. Blue squares represent 24 black holes discovered by X-ray method, while red circles represent 10 black holes and candidates by RV and astrometry methods. Open circles represent black hole candidates on debate, including MWC 656 and 2M05215658+4359220. The error bars represent 1$\sigma$ uncertainties. Objects with lower mass limits include GRS 1009-45, GS 1354-64, GC NGC3201 \#12560 and \#21859, VFTS 243, and AS 386.
    Panel e: orbital eccentricity versus orbital period.
    }
    \label{newbhs.fig}
\end{figure*}

\end{document}